\documentclass[a4paper,11pt]{article}
\pdfoutput=1
\usepackage{jheppub} % for details on the use of the package, please
\usepackage{graphicx,color,rotating}
\usepackage{hyperref}
\usepackage{epsfig,color}
\usepackage{slashed}
\usepackage{amsfonts}
\usepackage{ulem}
\usepackage{color}
\newcommand{\lsim}{\raisebox{-0.13cm}{~\shortstack{$<$ \\[-0.07cm] $\sim$}}~}
\newcommand{\gsim}{\raisebox{-0.13cm}{~\shortstack{$>$ \\[-0.07cm] $\sim$}}~}
\graphicspath{{D:/Desktop/draft}}
\usepackage{tabularx}
\usepackage{graphicx}
\usepackage{adjustbox}
\usepackage{tabularx}
\usepackage{subfig}

\def\lsim{\:\raisebox{-0.5ex}{$\stackrel{\textstyle<}{\sim}$}\:}
\def\gsim{\:\raisebox{-0.5ex}{$\stackrel{\textstyle>}{\sim}$}\:}

\begin{document}
\title{Phenomenology of bubble size distributions in a first-order phase transition}

\renewcommand{\thefootnote}{\arabic{footnote}}

\author{
Danny Marfatia$^{1}$, Po-Yan Tseng$^{2,3}$ and Yu-Min Yeh$^{2}$}
\affiliation{
$^1$ Department of Physics \& Astronomy, University of Hawaii at Manoa,
2505 Correa Rd., Honolulu, HI 96822, USA \\
$^2$ Department of Physics and CTC, National Tsing Hua University,
Hsinchu 300, Taiwan \\
$^3$ Physics Division, National Center for Theoretical Sciences,
Taipei 106319, Taiwan \\
}

%\pacs{14.80.Bn.,14.80.Da,14.80.Ec}
\date{\today}

\abstract{
In a cosmological first-order phase transition (FOPT), the true and false vacuum bubble radius distributions are not expected to be monochromatic, as is usually assumed. 
Consequently, Fermi balls (FBs) and primordial black holes (PBHs)
produced in a dark FOPT will
have extended mass distributions.
We show how gravitational wave~(GW), microlensing and Hawking evaporation signals for extended bubble radius/mass distributions deviate from the case of monochromatic distributions. 
The peak of the GW spectrum is shifted to lower frequencies, and the spectrum is broadened at frequencies below the peak frequency. Thus, the radius distribution of true vacuum bubbles
introduces another uncertainty in the evaluation of the GW spectrum from a FOPT. 
The extragalactic gamma-ray signal at AMEGO-X/e-ASTROGAM from PBH evaporation may evince a break in the power-law spectrum between 5~MeV and 10~MeV for an extended PBH mass distribution.
Optical microlensing surveys may observe PBH mass distributions with average masses below $10^{-10}M_\odot$, which is not possible for monochromatic mass distributions. This expands the FOPT parameter space that can be explored with microlensing.

}

\maketitle

\section{Introduction}

A novel scenario of sub-solar mass primordial black hole (PBH) formation via a first-order phase transition (FOPT) in a dark sector has been proposed~\cite{Gross:2021qgx,Kawana:2021tde}.
During the FOPT, the dark Dirac fermions in the false vacuum (FV) aggregate and get compressed to produce  macroscopic states called Fermi balls (FBs). The FBs may then subsequently collapse to produce PBHs. 
The conditions under which this collapse occurs depend on the strength of the Yukawa interaction and the highly model-dependent FB cooling rate. 

FB/PBH formation through a FOPT provides various phenomenological signals.
For example, PBHs produced from a FOPT at the MeV energy scale can be lighter than $10^{-16} M_\odot$ and copiously emit light particles through Hawking radiation, thus contributing to dark matter and cosmic ray fluxes in the present epoch~\cite{Marfatia:2021hcp,Marfatia:2022jiz,Kim:2023ixo}.
Alternatively, if FBs don't collapse to PBHs, or the PBHs are sufficiently long-lived, gravitational microlensing offers an indirect signal of the FOPT~\cite{Marfatia:2021twj}.
Since PBHs are point-like lenses, and FBs may behave as extended lenses, microlensing signatures of PBHs and FBs may be distinguishable. 

Previous phenomenological work has focused on monochromatic FB and PBH mass distributions from a FOPT assuming vacuum bubbles of a fixed size~\cite{Hong:2020est}. However, more generally, extended mass distributions can originate  
from the volume distribution of FV bubbles at the percolation temperature~\cite{Lu:2022paj}. 
The FV bubble radius distribution is obtained by calculating the probability of bubble walls collapsing to a point at which a FB forms. The probability depends on the distances between the bubble walls and the collision point at the percolation temperature, and thus on the volume of the FV bubble. Since Hawking evaporation is directly determined by the PBH mass, the diffuse cosmic ray fluxes may carry the imprint of the FV bubble distribution.

In this work, we investigate how the true vacuum (TV) bubble radius distribution modifies the GW spectrum from the double broken power law (DBPL) spectrum of Ref.~\cite{Caprini:2024hue}. 
We assess the ability of gravitational microlensing surveys and
extragalactic gamma-ray spectra to differentiate between an extended FB/PBH mass distribution and a monochromatic distribution within the FOPT framework.

The paper is organized as follows. In Section~\ref{sec:PBH_FOPT}, we review how the FV bubble radius distribution is obtained in a FOPT, and how this leads to extended FB/PBH mass distributions.
In Section~\ref{sec:GW}, we study how the TV bubble radius distribution modifies the GW spectrum.
Modifications to gravitational microlensing and PBH evaporation signals due to extended mass distributions are studied in Sections~\ref{sec:microlensing} and~\ref{sec:gamma_ray}, respectively. 
We summarize in Section~\ref{sec:summary}.

\bigskip

\section{Formation of Fermi balls and primordial black holes}
\label{sec:PBH_FOPT}

During a FOPT governed by a dark scalar $\phi$, dark Dirac fermions $\chi$'s get trapped in FV bubbles and form macroscopic nontopological solitons, i.e., Fermi balls. Once a FB cools sufficiently, the negative Yukawa potential energy may dominate the total FB energy, and cause the FB to collapse to a PBH.
Note that the criterion for FB collapse to a PBH relies on the cooling rate which is model dependent.
We sidestep this issue by supposing that FV bubbles evolve only into stable FBs or into PBHs. 

The above scenario can be realized by the minimal particle-level Lagrangian,
\begin{eqnarray}
\mathcal{L}\supset \bar{\chi} 
\left(i \slashed{\partial}-m \right) \chi 
-g_\chi \phi \bar{\chi} \chi -V_{\rm eff}(\phi,T) \,,
\end{eqnarray}
with the finite-temperature quartic  potential~\cite{Dine:1992wr,Adams:1993zs},
\begin{eqnarray}
\label{eq:Veff}
V_{\rm eff}(\phi,T)= D(T^2-T^2_0)\phi^2-(AT+C)\phi^3+\frac{\lambda}{4}\phi^4\,.
\end{eqnarray}
We chose to swap $T_0$ for the difference in energy density $B \equiv |V_{\rm eff}(\tilde\phi,0)|$ between the false vacuum and the true vacuum $\tilde\phi$ at zero temperature~\cite{Marfatia:2021hcp}.
So the input parameters that define the FOPT are 
$\lambda,A,B,C,D,m,g_\chi$.
The quartic potential determines the bubble nucleation rate per unit volume, $\Gamma(T)$, and the fraction of space in the FV, $F(t)$. We identify the temperature of the phase transition with the percolation temperature $T_\star$ when the volume fraction of space remaining in the false vacuum is $1/e$, i.e., $F(t_\star)=1/e \approx 0.37$. Assuming that individual false vacuum bubbles do not separate into smaller bubbles, the FV bubble radius~\cite{Hong:2020est,Kawana:2021tde} and the TV bubble radius~\cite{Ellis:2018mja} are respectively,
\begin{equation}
\label{eq:Rmono}
    {R_f}_\star \simeq \left( \frac{3 v_w}{4 \pi \Gamma_\star} \right)^{1/4}\,,~~~~ R_\star \simeq \pi^{1/3} \frac{v_w}{\beta} \simeq  \left( \frac{\pi^{1/3}v_w}{8\Gamma_\star} \right)^{1/4}\,,
\end{equation}
where $v_w$ is the bubble wall velocity, $\Gamma_\star \equiv \Gamma(T_\star)$, and $\beta^{-1}$ is the duration of the phase transition. 
We have identified the mean separation between TV bubble centers as the diameter of a TV bubble at percolation since there is a connected path between bubbles across the space at percolation.
%$\bar{R}_\star$ is a measure of the TV bubble radius. 
 The number density of FBs is given by $n_{\rm FB}|_{T_\star}V_\star = F(t_\star)$ if each spherical FV bubble of critical volume $V_\star = (4\pi/3){R^3_f}_{\star}$ results in one FB.
In terms of the asymmetry in the number densities 
$\eta_\chi \equiv (n_\chi - n_{\bar{\chi}})/s$, (with entropy density $s=(2 \pi^2/45)\left( g^{\rm SM}_{*s}T^3_{\rm SM}+g^{\rm D}_{*s}T^3 \right))$, the total number of dark fermions in a critical volume is given by 
\begin{equation}
\label{eq:QFB}
Q_{\rm FB}=\eta_\chi \left(\frac{s}{n_{\rm FB}} \right)_{T_\star}=\frac{\eta_\chi\, s}{F(t_\star)}
\left( \frac{4\pi}{3} {R^3_f}_\star \right)\,.
\end{equation}

A FV bubble can be considered as a Fermi gas of $Q_{\rm FB}$ particles coupled via a Yukawa interaction to a scalar with vacuum energy $V_0(T)\equiv V_{\rm eff}(0,T)-V_{\rm eff}(v_\phi(T),T)$. The stable configuration of this system is a FB. Complete expressions for the FB radius $R_{\rm FB}$ and mass $M_{\rm FB}$ can be found in Ref.~\cite{Marfatia:2021hcp}.
In the limit $V_0 \gg T^4, m^4$ and neglecting the Yukawa energy,
\begin{eqnarray}
\label{eq:FB_radius_mass}
    R_{\rm FB} \simeq Q^{1/3}_{\rm FB} 
    \left[ \frac{3}{16} \left( \frac{3}{2\pi} \right)^{2/3} \frac{1}{V_0} \right]^{1/4}\,, ~~~~
    M_{\rm FB} \simeq Q_{\rm FB} (12 \pi^2 V_0)^{1/4}\,,
\end{eqnarray}
This monochromatic mass distribution is obtained by using the average radius ${R_f}_\star$ of false vacuum bubbles. 
In the next subsection we generalize this result by using 
an extended FV radius distribution.

The $\chi$'s inside a FB are coupled to $\phi$ by the attractive Yukawa interaction $g_\chi \phi \bar{\chi} \chi$. Since the interaction length $L_\phi = [2D(T^2-T^2_0)]^{-1/2}$ is temperature dependent,
as the FB cools, at temperature $T_\phi$, the $L_\phi$ becomes comparable to the mean separation distance of $\chi$'s inside the FB. If the negative Yukawa energy dominates the FB total energy, the FB collapses to a PBH. We are interested in FOPTs for which $T_\star > T_\phi$, in which case a FB is formed first, and a FV bubble does not collapse directly to a PBH~\cite{Marfatia:2021hcp}.
Then, the PBH mass at formation can be identified as the FB mass evaluated at $T_\phi$, i.e. $M_{\rm PBH}=M_{\rm FB}(T_\phi)$.

\bigskip

\subsection{Extended mass distributions}

We are interested in FOPTs whose duration $\beta^{-1}$ is much shorter than the Hubble time $H_\star^{-1}\equiv H(t_\star)^{-1}$. Then, the radius of a TV bubble at time $t$ that was nucleated at time $t'$ is $R(t,t')=v_w(t-t')$.
Neglecting the evolution of the scale factor $a(t)$, the fraction of space in the FV at time $t$ is
\begin{equation}
    F(t)=e^{-I(t)}\,,~~~~
    I(t)\equiv \int^{t}_{t_c} dt' \Gamma(t')\frac{4\pi}{3}R^3(t,t')\,, 
\end{equation}
where $t_c$ is the time when the critical temperature $T_c$ is reached.
The radius distribution of TV bubbles at time $t$ is determined by the average nucleation rate at the earlier time $t'=t-R/v_w$ and is given by~\cite{Lu:2022paj}
\begin{equation}
\label{eq:R_distribution}
    \frac{dn_{\rm TV}}{dR}(t)=\frac{1}{v_w}F(t')\Gamma(t')\,.
\end{equation}
Analogously, by considering a reverse time description with $t=t'-R_f/v_w$, FV bubbles (of radius $R_f$) have the distribution,
\begin{equation}
    \frac{dn_{\rm FV}}{dR_f}(t)=\frac{1}{v_w}\left(1- F(t')\right)\Gamma_f(t')\,,
\end{equation}
where $\Gamma_f$ is the FV nucleation rate.

Since the minimum number of TV bubbles needed to enclose a FV volume is four, a FV bubble can be approximated as a tetrahedron.
    Using $\Gamma(t)\simeq \Gamma_\star e^{\beta(t-t_\star)}$ since $\beta/H_\star \gg 1$, the saddle point approximation $I(t)\simeq I_\star e^{\beta(t-t_\star)}$ with $I_\star=8\pi v^3_w \Gamma_\star/\beta^4 =-\ln(F(t_\star)) =1$,
and the tetrahedral approximation for the shape of the FV bubbles (which gives $\Gamma_f(t) \simeq {\frac{I_\star^4\beta^4}{192v_w^3}}e^{4\beta (t-t_\star)}e^{-I(t)}$), the FV bubble radius distribution at percolation is~\cite{Lu:2022paj}\footnote{The expression in Ref.~\cite{Lu:2022paj} is missing a factor of $1/v_w$ in the normalization.} 
\begin{eqnarray}
\label{eq:Rr_distribution}
	\frac{dn_{\rm FV}}{dR_f}(t_\star)\simeq\frac{ \beta^4}{192v^4_w}e^{\frac{4\beta R_f}{v_w}}e^{-e^{\frac{\beta R_f}{v_w}}}
	\left(1-e^{-e^{\frac{\beta R_f}{v_w}}}\right)\,.
\end{eqnarray}
 Note that the monochromatic FV bubble radius in Eq.~(\ref{eq:Rmono}) can be expressed as ${R_f}_\star \simeq 6^{1/4} v_w/\beta$~\cite{Hong:2020est,Kawana:2021tde}. 
Similarly, the radius distribution of TV bubbles is
\begin{eqnarray}
\label{eq:R_distribution}
	\frac{dn_{\rm TV}}{dR}(t_\star)\simeq\frac{ \beta^4}{8\pi v^4_w} e^{-\frac{\beta R}{v_w}} e^{-e^{-\frac{\beta R}{v_w}}} \,.
\end{eqnarray}

\begin{figure}[t]
        \centering
        {\includegraphics[width=9cm,angle=270]{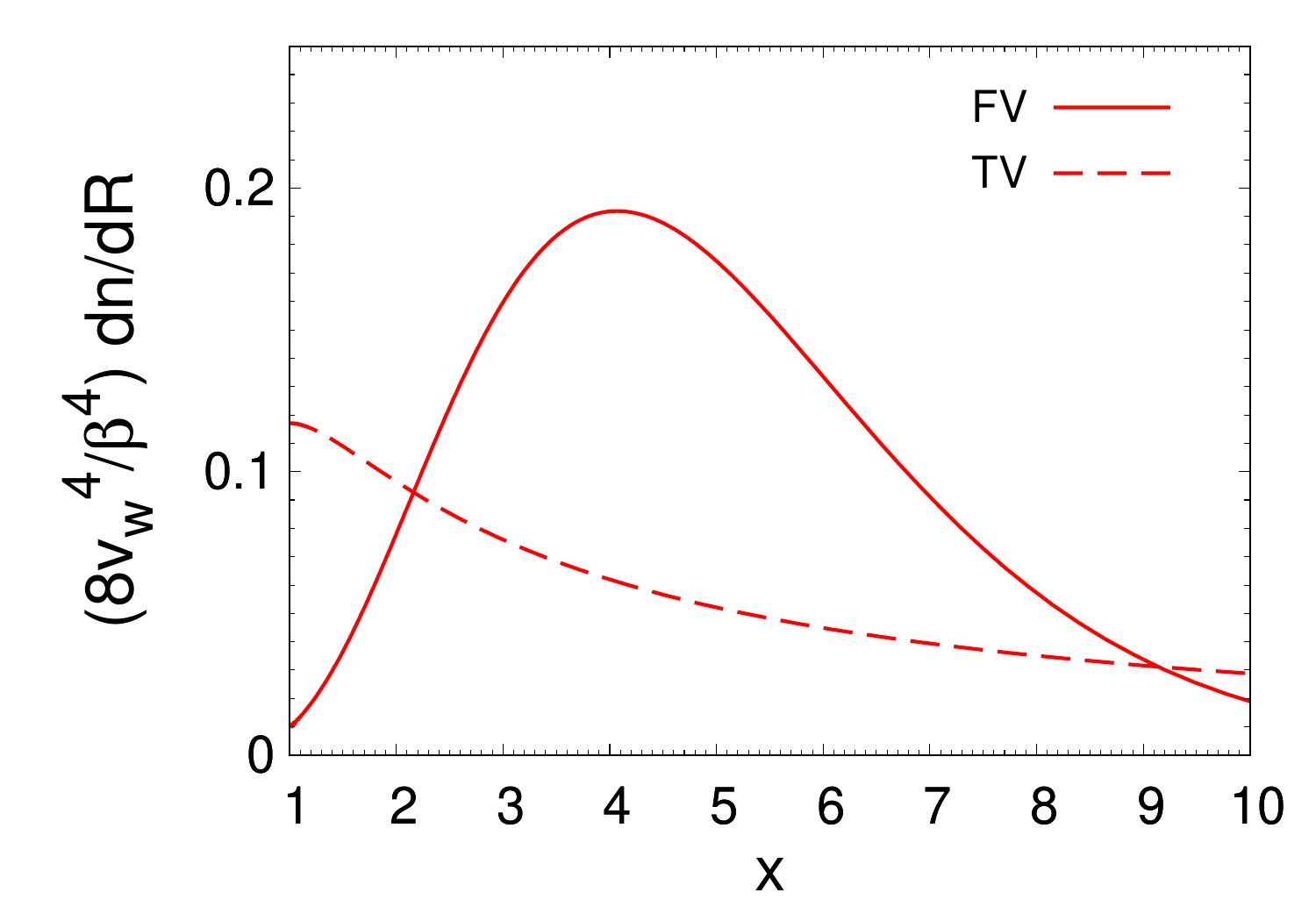}}
        \caption{Comparison between the FV and TV radius distributions in Eq.~(\ref{eq:x}). 
        }\label{fig:dndR_TV_FV}
\end{figure}

It is convenient to write Eqs.~(\ref{eq:Rr_distribution}) and~(\ref{eq:R_distribution}) in terms of $x \equiv \exp\left(\beta R_{(f)}/v_w \right)>1$:
	\begin{equation}
	\frac{dn_{\rm FV}}{dR_f}(t_\star)\simeq\frac{\beta^4}{192v^4_w}x^4 e^{-x}(1-e^{-x})\,,~~~~ \frac{dn_{\rm TV}}{dR}(t_\star)\simeq\frac{\beta^4}{8\pi v^4_w} {1\over x} e^{-{1\over x}}\,.
 \label{eq:x}
	\end{equation}
 In Fig.~\ref{fig:dndR_TV_FV}, we compare the FV and TV radius distributions. We find the average radii at percolation to be $\langle R_f\rangle_\star= 1.32 v_w/\beta$ 
 and $\langle R\rangle_\star= 1.26 v_w/\beta$. 

\begin{figure}[t]
        \centering
        {\includegraphics[width=9cm,angle=270]{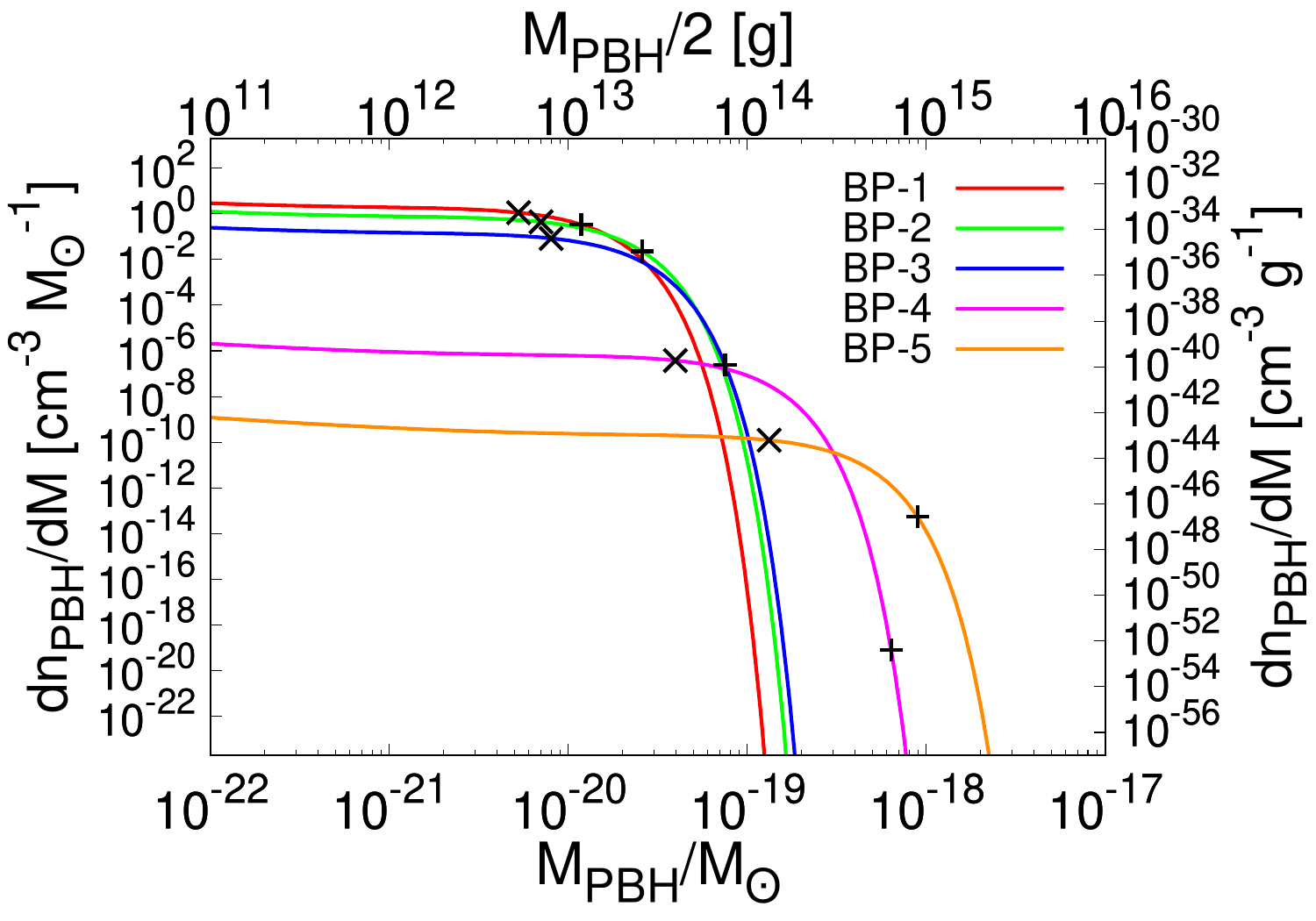}}
        \caption{The extended mass distributions (at PBH production) for the benchmark points in Table~\ref{table1}. The crosses mark the average mass; the pluses mark the corresponding monochromatic mass. 
        }\label{figmd}
\end{figure}

Because the FV bubbles are not truly spherical, the total number of $\chi$'s in Eq.~(\ref{eq:QFB}) is obtained  by writing  the volume of a FV bubble as $\bar{A}\frac{4\pi}{3} R^3_f$:
\begin{eqnarray}
\label{eq:QFB_tetrohedron}
Q_{\rm FB}=\frac{\eta_\chi s(t_\star)}{F(t_\star)}\bar{A} \frac{4\pi}{3} R^3_f\,,~~~~
{F(t_\star)} = \bar{A} \int dR_f \frac{4\pi R^3_f}{3} \frac{dn_{\rm FV}}{dR_f}(t_\star)={1\over e}\,.
\end{eqnarray}
The second equation yields $\bar{A}=0.96$ by normalizing to $F(t_\star)$. That $\bar{A}$ is close to unity, shows the validity of the tetrahedral approximation. 
Since $M_{\rm PBH}\simeq M_{\rm FB}\propto Q_{\rm FB} \propto R_f^3$,
we trade $M_{\rm FB/PBH}$ for $R_f$ 
and obtain the normalized mass distribution functions for FBs/PBHs:
\begin{eqnarray}
{\text f}(M_{\rm FB/PBH})\equiv \frac{\frac{4\pi R^3_f}{3}\, \frac{dn_{\rm FB/PBH}}{dM_{\rm FB/PBH}}}{\int dM_{\rm FB/PBH}\, \frac{4\pi R^3_f}{3}\, \frac{dn_{\rm FB/PBH}}{dM_{\rm FB/PBH}}}\,,
\end{eqnarray}
where 
\begin{equation}
\frac{dn_{\rm FB/PBH}}{dM_{\rm FB/PBH}}= \frac{R_f}{3M_{\rm FB/PBH}} \frac{dn_{\rm FV}}{dR_f}(t_\star)\,.
\label{dist}
\end{equation}
From these mass distributions, the average FB/PBH mass $\langle M_{\rm FB/PBH} \rangle$ can be found. The extended mass distributions, the average mass and the monochromatic mass corresponding to the FOPT are shown in Fig.~\ref{figmd} for the benchmark points in Table~\ref{table1}.

\begin{table}[t]
        \centering
        \resizebox{\textwidth}{!}{
        \begin{tabular}{c|c c c c c c c}
            \hline
            \hline
            & \bf{BP-1}  & \bf{BP-2} & \bf{BP-3} & \bf{BP-4}  & \bf{BP-5} & \bf{BP-6} & \bf{BP-7} \\
           \hline
           \hline
            $B^{1/4}/{\rm MeV}$ & 3.803  & 4.034 & 2.243 & 0.365  & 0.204 & 19.53 & 0.0444 \\
            $\lambda$ & 0.0796  & 0.191 & 0.180 & 0.184  & 0.115 & 0.160 & 0.154\\
            $D$ & 0.640  & 0.556 & 0.365 & 0.282 & 0.385 & 1.828 & 0.224\\
            $\eta_{\chi}$ & $1.021\times10^{-15}$  & $3.977\times10^{-16}$ & $3.997\times10^{-16}$ & $3.329\times10^{-17}$  & $2.676\times10^{-18}$ & $7.293\times10^{-10}$ & $1.115\times10^{-9}$ \\
            $T_\star/T_{\rm SM\star}$ & 0.312  & 0.314 & 0.410 & 0.466  & 0.349 & 0.166 & 0.393\\
            $C/{\rm MeV}$ & 0.0639  & 0.518 & 0.251  & 0.0490  & 0.0196 & 2.784 & $6.274\times 10^{-3}$\\
            $g_\chi$ & 0.340  & 0.683 & 1.372  & 1.739  & 0.535 & 1.226 & 1.752 \\
            $m/B^{1/4}$ & 0.0776  & 0.324 & 0.678 & 0.138  & 0.185 & 0.201 & 0.129 \\
            \hline
            $\alpha$ & $3.610\times10^{-2}$ & $2.510\times10^{-2}$ & $2.863\times10^{-2}$ & $3.138\times10^{-2}$ & $3.498\times10^{-2}$ & $4.983\times10^{-2}$ & $1.836\times10^{-2}$ \\
            $\beta/H_\star$ & $5.091\times 10^3$ & $3.088\times 10^3$ & $3.271\times 10^3$ & $2.456\times 10^3$ & $1.530\times 10^3$ & $1.140\times 10^3$ & $1.505\times10^3$\\
            $T_\star/{\rm MeV}$ & 2.621 & 3.016 & 2.252 & 0.435 & 0.182 & 5.771 & 0.0640\\
            ${R_f}_\star/{\rm GeV^{-1}}$ & $3.151\times 10^{19}$ & $4.672\times 10^{19}$ & $1.334\times 10^{20}$ & $6.433\times 10^{21}$ & $3.910\times 10^{22}$ & $1.088\times 10^{19}$ & $5.086\times 10^{23}$\\
            $\langle{R_f}\rangle_\star/{\rm GeV^{-1}}$ & $4.098\times 10^{19}$ & $5.133\times 10^{19}$ & $1.447\times 10^{20}$ & $6.629\times 10^{21}$ & $3.633\times 10^{22}$ & $9.781\times 10^{18}$ & $4.412\times 10^{23}$ \\
            ${R}_\star/{\rm GeV^{-1}}$ & $2.949\times 10^{19}$ & $4.372\times 10^{19}$ & $1.248\times 10^{20}$ & $6.020\times 10^{21}$ & $3.659\times 10^{22}$ & $1.018\times 10^{19}$ & $4.759\times 10^{23}$ \\
            $\langle{R}\rangle_\star/{\rm GeV^{-1}}$ & $5.959\times 10^{19}$ & $8.389\times 10^{19}$ & $2.317\times 10^{20}$ & $1.034\times 10^{22}$ & $5.940\times 10^{22}$ & $1.629\times 10^{19}$ & $6.806\times 10^{23}$ \\
            $M_{\rm PBH}$/$M_{\odot}$ & $1.185\times 10^{-20}$  & $2.592\times 10^{-20}$ & $7.553\times 10^{-20}$ & $6.399\times 10^{-19}$  & $8.952\times 10^{-19}$ & - & -\\
            $\langle M_{\rm PBH}\rangle$/$M_{\odot}$ & $5.289\times 10^{-21}$  & $7.065\times 10^{-21}$ & $8.048\times 10^{-21}$ & $3.981\times 10^{-20}$  & $1.331\times 10^{-19}$ & - & - \\
            $M_{\rm FB}$/$M_{\odot}$ & -  & - & - & -  & - & $1.273\times 10^{-13}$ & $5.346\times 10^{-9}$\\
            $\langle M_{\rm FB}\rangle$/$M_{\odot}$ & -  & - & - & - & - & $5.925\times 10^{-12}$ & $2.335\times 10^{-8}$ \\
            $f_{{\rm PBH}/{\rm FB}}$ & $3.114 \times 10^{-8}$  & $1.392\times 10^{-8}$ & $9.318\times 10^{-9}$  & $1.466\times 10^{-10}$  & $5.512\times 10^{-12}$ & 0.0944 & $7.371\times10^{-4}$ \\
            $\Delta N_{\rm eff}$ & 0.250 & 0.181 & 0.253 & 0.326 & 0.266 & 0.292 & 0.254 \\
            \hline
            \hline
        \end{tabular} 
        }
        \centering
        \caption{
        Benchmark points with $A=0.1$.
        ${R_f}_\star$ and $R_\star$ are the radii of the monochromatic FV and TV bubble distributions in Eq.~(\ref{eq:Rmono}).
        $\langle{R_f}\rangle_\star$ and $\langle{R}\rangle_\star$ are the average radii of FV and TV bubbles at percolation. {\bf BP-6} and {\bf BP-7} produce FBs that do not collapse to PBHs.
        }
        \label{table1}
    \end{table}

\section{Gravitational wave spectrum from extended bubble radius distribution}
\label{sec:GW}

During the FOPT, latent heat is transferred to the bulk motion of the fluid. The compression component of the fluid motion propagates and sources GWs. Lattice simulations of the coupled scalar-fluid system show that the bubble size at percolation imprints itself on the GW spectra~\cite{Hindmarsh:2013xza,Cutting:2019zws}.
The GW spectrum due to sound waves in the plasma associated with TV bubbles of average radius $R$ at $t_\star$ can be modeled as a double broken power law (DBPL) with spectral breaks at frequencies $f_1$ and $f_2$~\cite{Caprini:2024hue},
\begin{eqnarray}
\label{eq:GW}
    \Omega^{\rm DBPL}_{\rm GW}(f,R)&=&\Omega_2 \times S_2(f)\,, \\
    S(f)&=&N \left( \frac{f}{f_1} \right)^{3} \left[1+\left( \frac{f}{f_1} \right)^{2} \right]^{-1}
                                            \left[1+\left( \frac{f}{f_2} \right)^{4} \right]^{-1}\,,~~~~ S_2(f) \equiv \frac{S(f)}{S(f_2)}\,, \nonumber
\end{eqnarray}
where $\Omega_2$ is the amplitude of the spectrum at $f_2$ since $S_2(f_2)=1$.
The frequency breaks $f_1$ and $f_2$ depend on $R$ via
\begin{equation}
\label{eq:GW_f}
    f_1 \simeq 0.2 \frac{H_{\star,0}}{H_\star R}\,,~~~~f_2 \simeq 0.5 \frac{H_{\star,0}}{\Delta_w\, H_\star R}\,,
\end{equation}
where $\Delta_w \equiv (|v_w-c_s|)/{\rm max}(v_w,c_s)$ with the speed of sound $c_s=1/\sqrt{3}$. 
Here, 
$H_{\star,0}$ is the redshifted value of $H_\star$ today.
$\Omega_2$ is given by~\cite{Caprini:2024hue}
\begin{equation}
\label{eq:GW_O2}
    \Omega_2 \simeq \frac{0.11}{\pi} \left(\sqrt{2}+\frac{2f_2/f_1}{1+f^2_2/f^2_1} \right) \left( \frac{a_\star}{a_0} \right)^{4} \left( \frac{H_\star}{H_0} \right)^{2}  \left( H_\star \tau_{\rm sw} \right)\left( H_\star R \right) K^2\,,
\end{equation}
where the kinetic energy fraction,
$K\simeq 0.6 \alpha/(1+\alpha)$, with  $\alpha$ the ratio of the latent heat to the total radiation energy density of the dark and visible sectors at $T_\star$. 
The lifetime of the sound waves $\tau_{\rm sw}$ in units of the Hubble time $H_\star^{-1}$ is $H_\star\tau_{\rm sw}={\rm min}[2H_\star R /\sqrt{3K},1]$, the shorter of the decay time into turbulence or the Hubble time.

The template of Ref.~\cite{Caprini:2024hue} described above has some limitations. It employs an average bubble radius defined as $R=(8\pi)^{1/3}{\rm max}(v_w,c_s)/\beta$, which as noted in Ref.~\cite{Caprini:2019egz}, needs to be refined to obtain accurate predictions of the GW spectrum because sound shells of this radius carry most of the energy of the FOPT. 
Also, numerical simulations of the coupled scalar-fluid system are performed with bubbles nucleated simultaneously~\cite{Cutting:2019zws}, while bubble nucleation should occur at an exponentially increasing rate.  
Although in the sound shell model, an extended bubble radius distribution can be obtained {\it at collision} for both simultaneous and exponential nucleation~\cite{Hindmarsh:2019phv}, 
a rigorous analysis would entail large and long duration scalar-hydrodynamic simulations with exponential nucleation. Such simulations do not exist. 

With these caveats in mind, we adopt a phenomenological approach to illustrate the impact of an extended bubble radius distribution on the GW spectrum by applying the template to ensembles of vacuum bubbles with different average radii.
Absent knowledge of the range of $R$ for which the template is valid, a quantitative interpretation of our results should be made with caution. 

For a monochromatic bubble radius distribution, we substitute the average TV bubble radius at percolation, 
\begin{equation}
\langle R \rangle_\star = {\frac{\int dR \,R \frac{dn_{\rm TV}(t_\star)}{dR}}{\int dR\, \frac{dn_{\rm TV}(t_\star)}{dR}}}\,,
\label{rstar}
\end{equation}
into Eq.~(\ref{eq:GW}), i.e., $\Omega^{\rm mono}_{\rm GW}(f)=\Omega^{\rm DBPL}_{\rm GW}(f,\langle R \rangle_\star)$.
For an extended distribution, the GW spectrum is obtained by convolving Eq.~(\ref{eq:GW}) with Eq.~(\ref{eq:R_distribution}):
\begin{equation}
\label{eq:GW_dndR}
    \Omega_{\rm GW}(f) = \frac{\int dR\,  \frac{dn_{\rm TV}(t_\star)}{dR}\, \Omega^{\rm DBPL}_{\rm GW}(f,R)}{\int\, dR \, \frac{dn_{\rm TV}(t_\star)}{dR}}  \,.
\end{equation}

  \begin{figure}[t]
		\centering
        \includegraphics[width=9cm,angle=270]{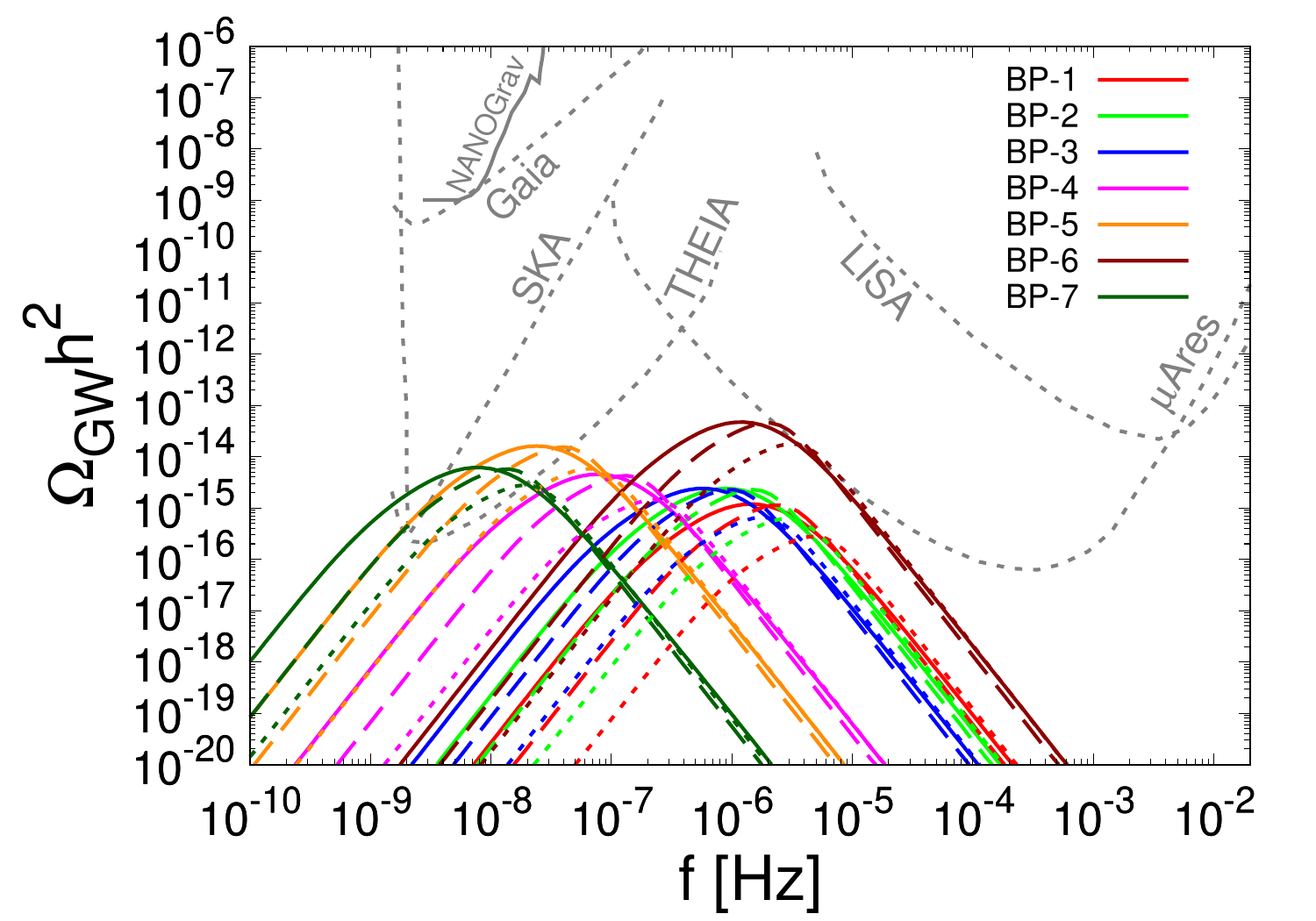}
		\caption{
  The solid curves are the GW spectra for the benchmark points in Table~\ref{table1} for extended bubble radius distributions. The dashed and dotted curves show the corresponding spectra for monochromatic radius distributions, evaluated at $\langle R \rangle_\star$ (Eq.~\ref{rstar}) and $R_\star$ (Eq.~\ref{eq:Rmono}), respectively. 
  }
		\label{fig:GW}
	\end{figure}

In Fig.~\ref{fig:GW}, the solid curves are the GW spectra for the BPs in Table~\ref{table1} for extended bubble radius distributions. The dashed and dotted curves  show the corresponding spectra for the monochromatic cases evaluated at $\langle R \rangle_\star$ (Eq.~\ref{rstar}) and $R_\star$ (Eq.~\ref{eq:Rmono}), respectively.
The dashed curves have a lower peak frequency and higher amplitude than the dotted curves because for bubbles of a given radius $R$, the amplitude 
$\Omega_2 \propto R \tau_{\rm sw} \sim R^2$, the peak frequency scales inversely with $R$ (see Fig.~1 of Ref.~\cite{Caprini:2024hue}), and $\langle R \rangle_\star > R_\star$ for all the {\bf BP}s.
Although on the scale of the figure, the amplitudes of the solid and dashed curves are almost the same, the amplitudes of the solid curves are higher than the dashed curves. 
 Also, the peak frequencies for the extended distributions shift below those for the monochromatic distributions.{\footnote{Interestingly, our approach reproduces the shift in peak frequency of the GW spectrum to lower values found in a simplified model of exponential nucleation; see Fig.~6 of Ref.~\cite{Hindmarsh:2019phv}.} 
While spectral broadening is expected for an extended distribution, we find that the broadening is far more significant for frequencies below the peak. 
This is because for an extended $dn_{\rm TV}/dR$ distribution,
the amplitude of GWs for bubbles with $R > \langle{R}\rangle_\star$ (corresponding to lower peak frequencies) is enhanced compared to bubbles with $R < \langle R\rangle_\star$, leading to the asymmetric spectral broadening.
Above the peak, the GW spectrum modeled by DBPL or by a single power law is essentially the same.
Clearly, the effect of the TV bubble radius distribution on the GW spectrum should be taken into account as per Eq.~(\ref{eq:GW_dndR}). In fact, the bubble radius distribution introduces another uncertainty in the evaluation of the GW spectrum.

\bigskip

\section{Gravitational microlensing}
\label{sec:microlensing}

Microlensing occurs when a massive object, e.g., FB or PBH, passes through the line of sight of a background star, causing the luminosity of the star to be enhanced and then restored to its original value. Transient brightening is the distinctive signature of microlensing. Subaru Hyper Suprime-Cam (HSC)~\cite{Niikura:2017zjd} has surveyed $8.7\times 10^7$ stars in the M31 galaxy, located 770~kpc from the center of the Milky Way.
During seven hours of observation, only one transient brightening event was identified. This result sets strong bounds on the mass and abundance of macroscopic dark matter candidates, including FBs and PBHs.

\bigskip

\subsection{Microlensing by Fermi balls}
\label{sec:microlensing_FB}
We consider FBs to be stable on timescales of the age of the Universe if the (negative) Yukawa energy does not dominate the total energy of the FB before the end of the FOPT (defined as the time at which 99\% of the space is filled by the true vacuum).
Microlensing due to monochromatic mass FBs has been discussed in Ref.~\cite{Marfatia:2021twj}. It was shown that FBs have a uniform density profile and do not necessarily behave like point-like lenses. We extend this work to investigate the impact of extended mass distributions.

To generalize to an extended mass distribution, we treat the differential event rate per source star as a function of $M_{\rm FB}$, i.e., $\frac{d^2\Gamma(M_{\rm FB})}{dx dt_E}$, where $t_E$ is the time for which the magnification is above threshold and $x$ is the ratio of the Earth-lens distance to the Earth-source distance; see Ref.~\cite{Marfatia:2021twj}. We adopt NFW dark matter halo profiles for M31 and the Milky Way, and the stellar radius distribution $dn/dR_S$ in M31 from Ref.~\cite{Smyth:2019whb}.
The FB mass distribution is related to the FV bubble distribution at percolation via Eq.~(\ref{dist}).
Then, the number of microlensing events expected at Subaru-HSC is
\begin{eqnarray}
N_{\rm events}= \frac{N_S T_{\rm obs}}{\int dM_{\rm FB} \frac{dn_{\rm FB}}{dM_{\rm FB}}} \int^{t_{\rm max}}_{t_{\rm min}} dt_E \int dR_S \int dM_{\rm FB}
\int^1_0 dx \frac{d^2 \Gamma (M_{\rm FB})}{dx dt_E} \frac{dn}{dR_S} \frac{dn_{\rm FB}}{dM_{\rm FB}}\,,
\end{eqnarray}
where $N_S=8.7\times 10^7$ is the number of stars in the survey, $T_{\rm obs}=7$~hours is the observational period, $t_{\rm min}=2$~minutes is the cadence, and $t_{\rm max}=7$~hours. Since only one transient event was observed, the 
95\%~C.L. upper limit corresponds to $N_{\rm events}=4.74$. In the future, with 70 hours of observation (with $t_{\rm max}$ set conservatively to 7~hours), we evaluate 
the 95\%~C.L. sensitivity by requiring $N_{\rm events}\le 16.96$.

We scan over six FOPT parameters with $A=0.1$ and adjust $\eta_\chi$ so that the relic abundance satisfies $\Omega_{\rm FB}h^2\,, \Omega_{\rm PBH}h^2 \le 0.12$. 
We require the effective number of extra neutrinos $\Delta N_{\rm eff}$ at the end of the phase transition to be smaller than 0.5.
We select the seven benchmark points (BPs) listed in Table~\ref{table1}.
Only {\bf BP-6} and {\bf BP-7} produce FBs that do not collapse to PBHs.

\begin{figure}[t]
      \centering
      \includegraphics[height=1.1in,angle=0]{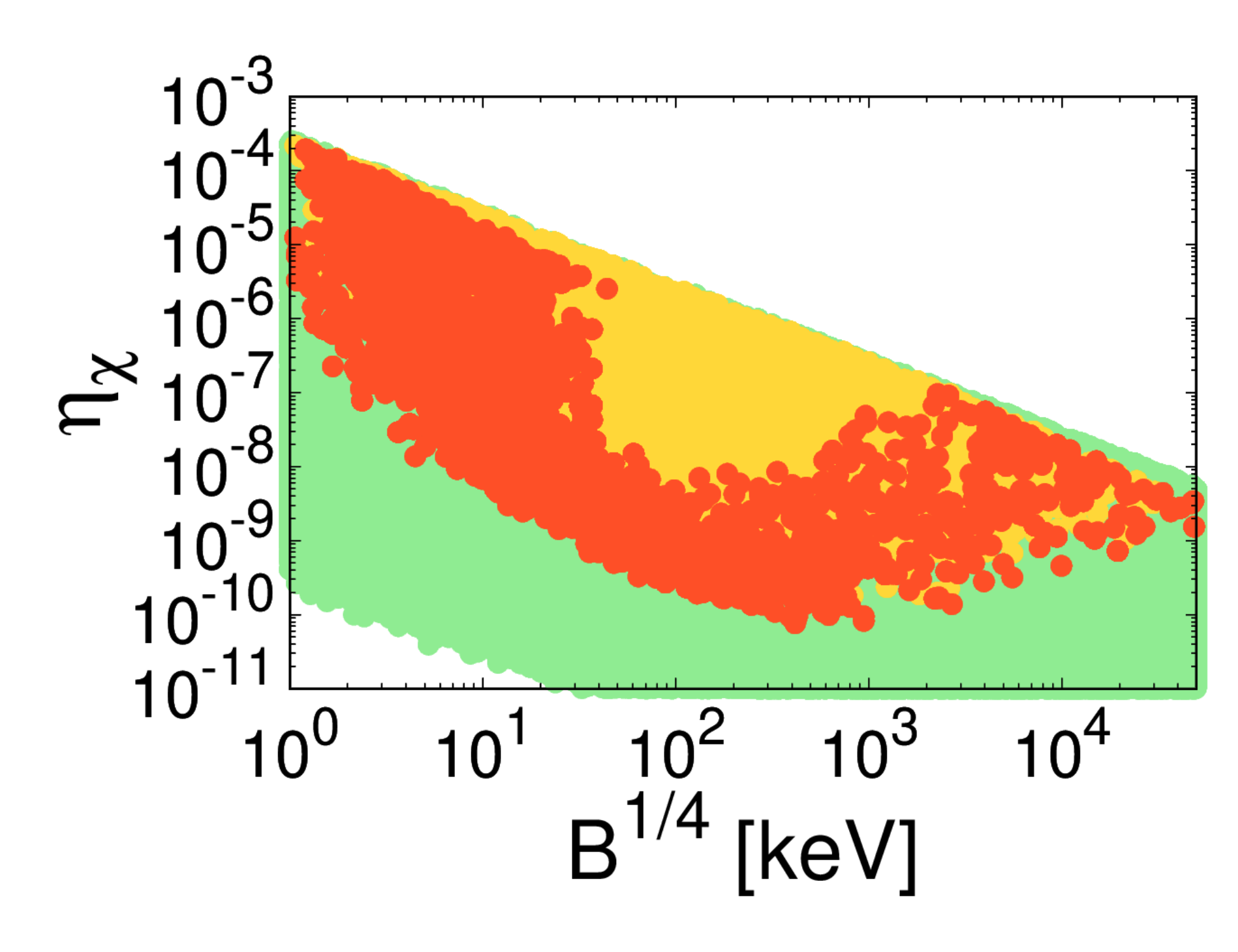}
      \includegraphics[height=1.05in,angle=0]{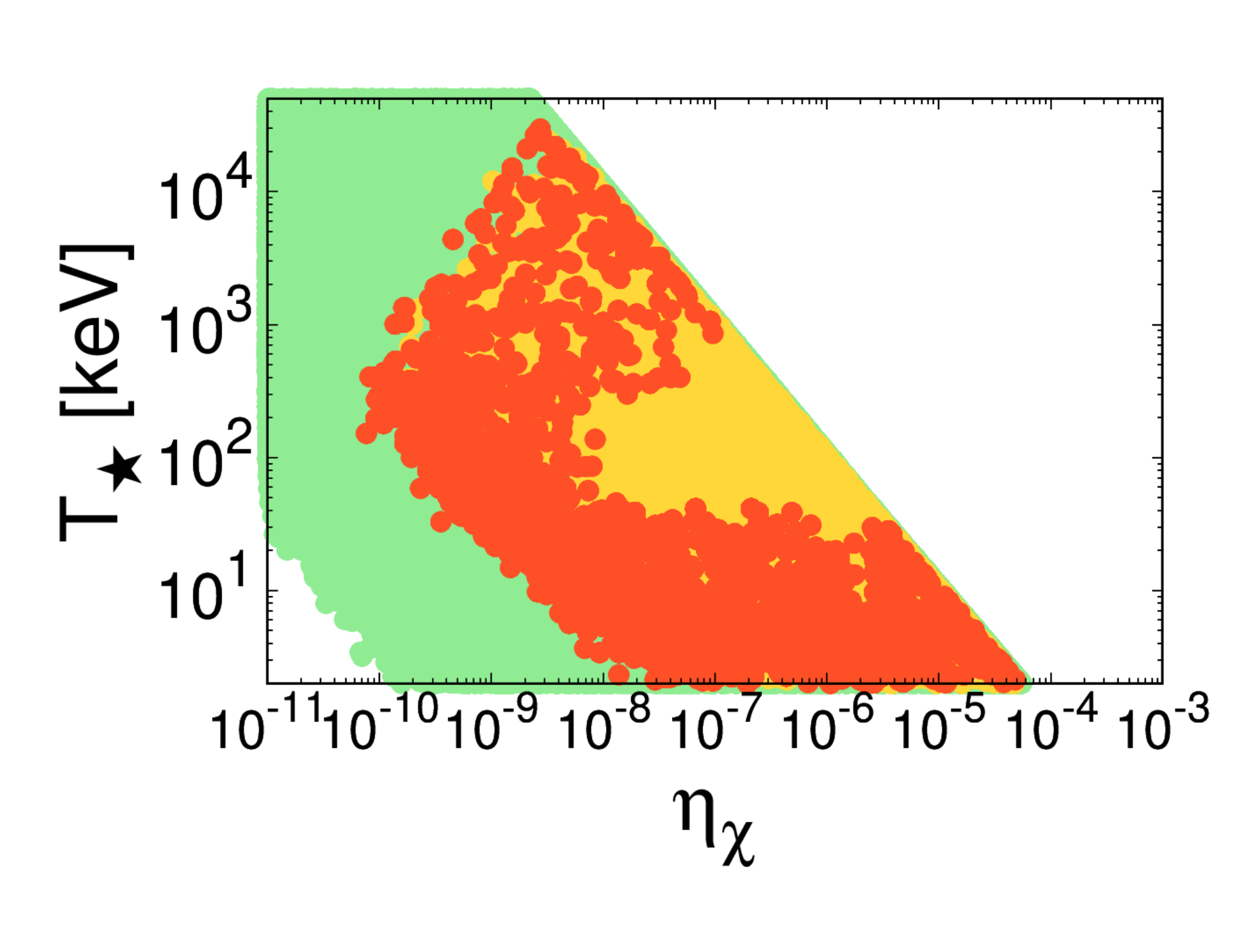}
      \includegraphics[height=1.1in,angle=0]{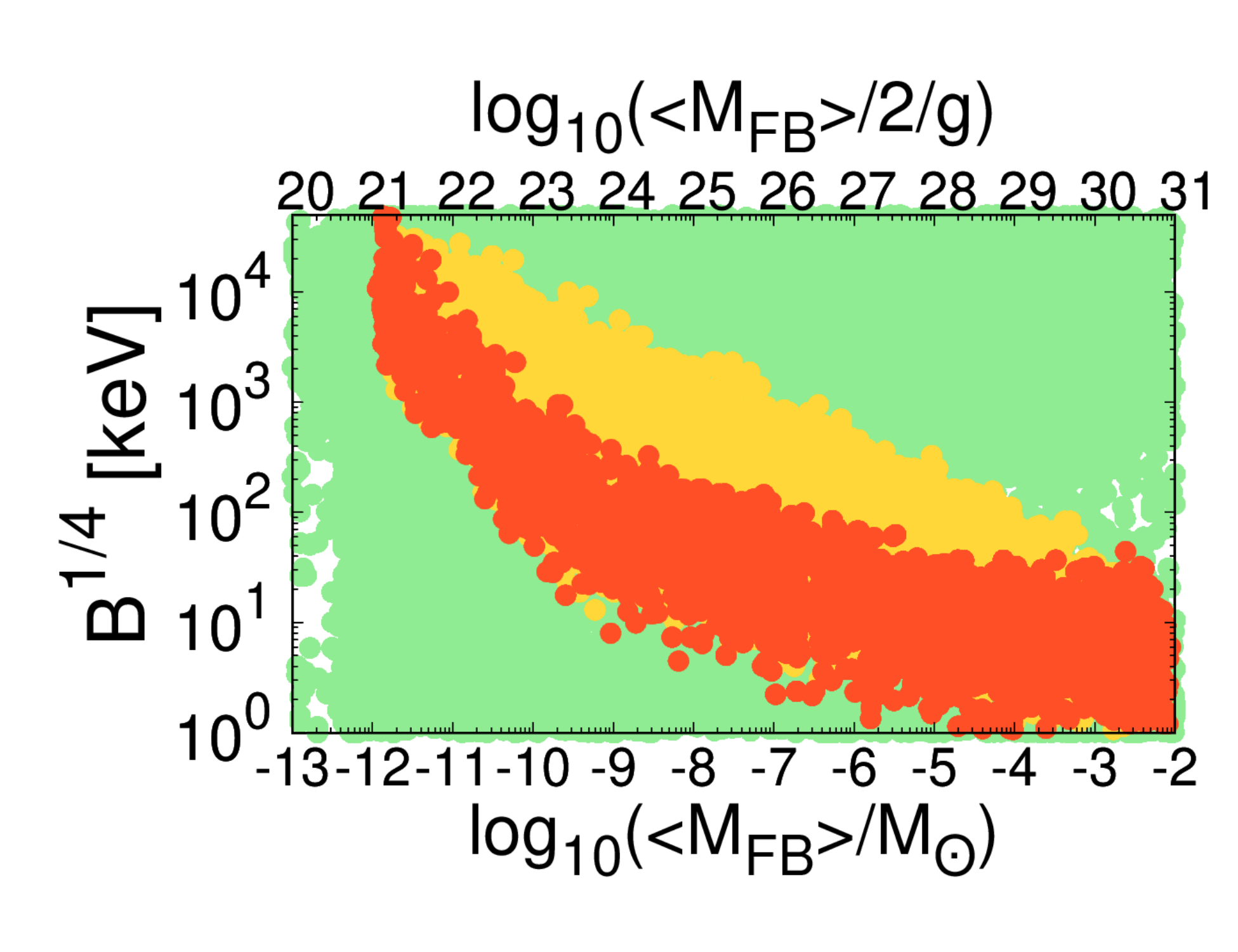}
      \includegraphics[height=1.1in,angle=0]{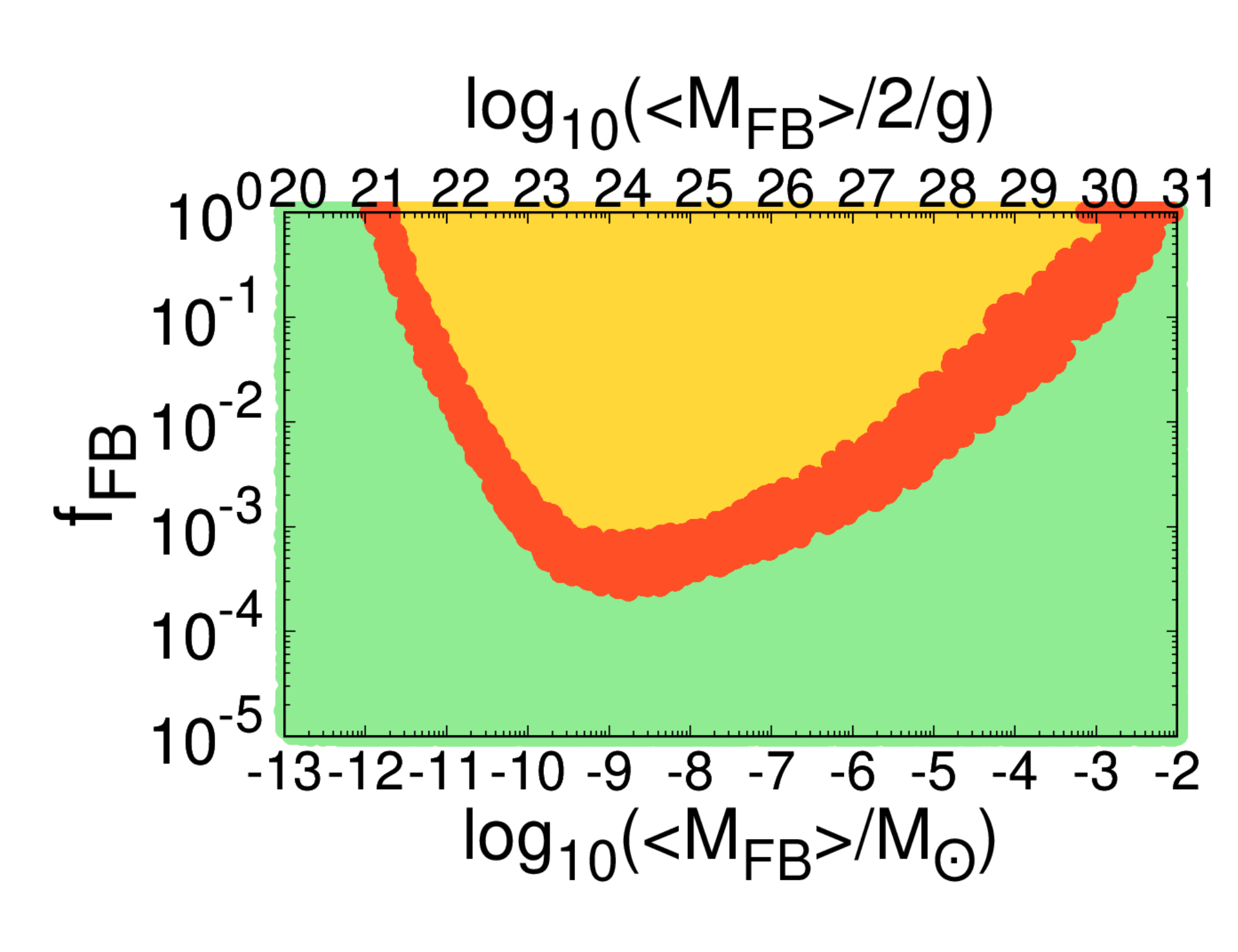}
      \includegraphics[height=1.1in,angle=0]{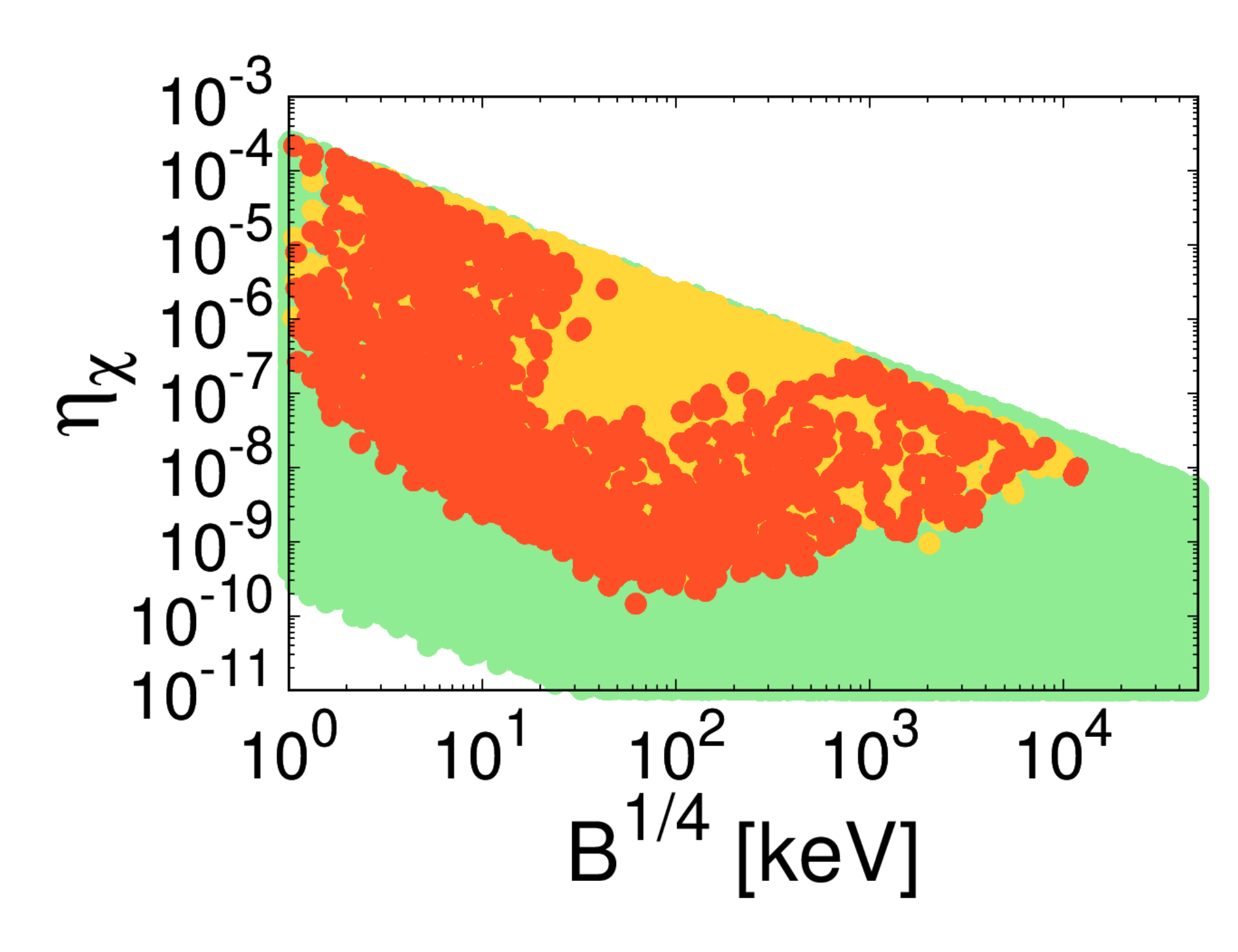}
      \includegraphics[height=1.05in,angle=0]{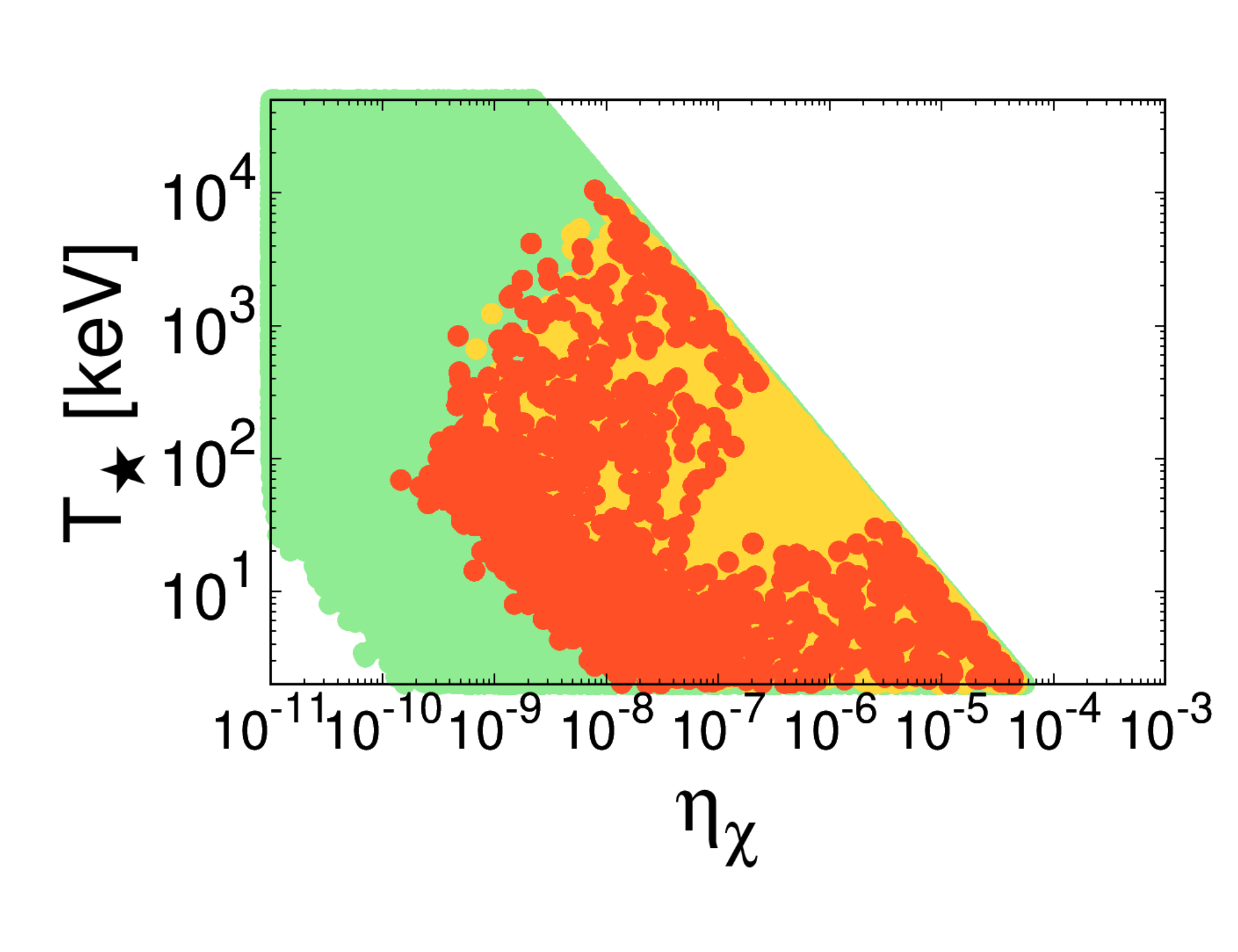}
      \includegraphics[height=1.1in,angle=0]{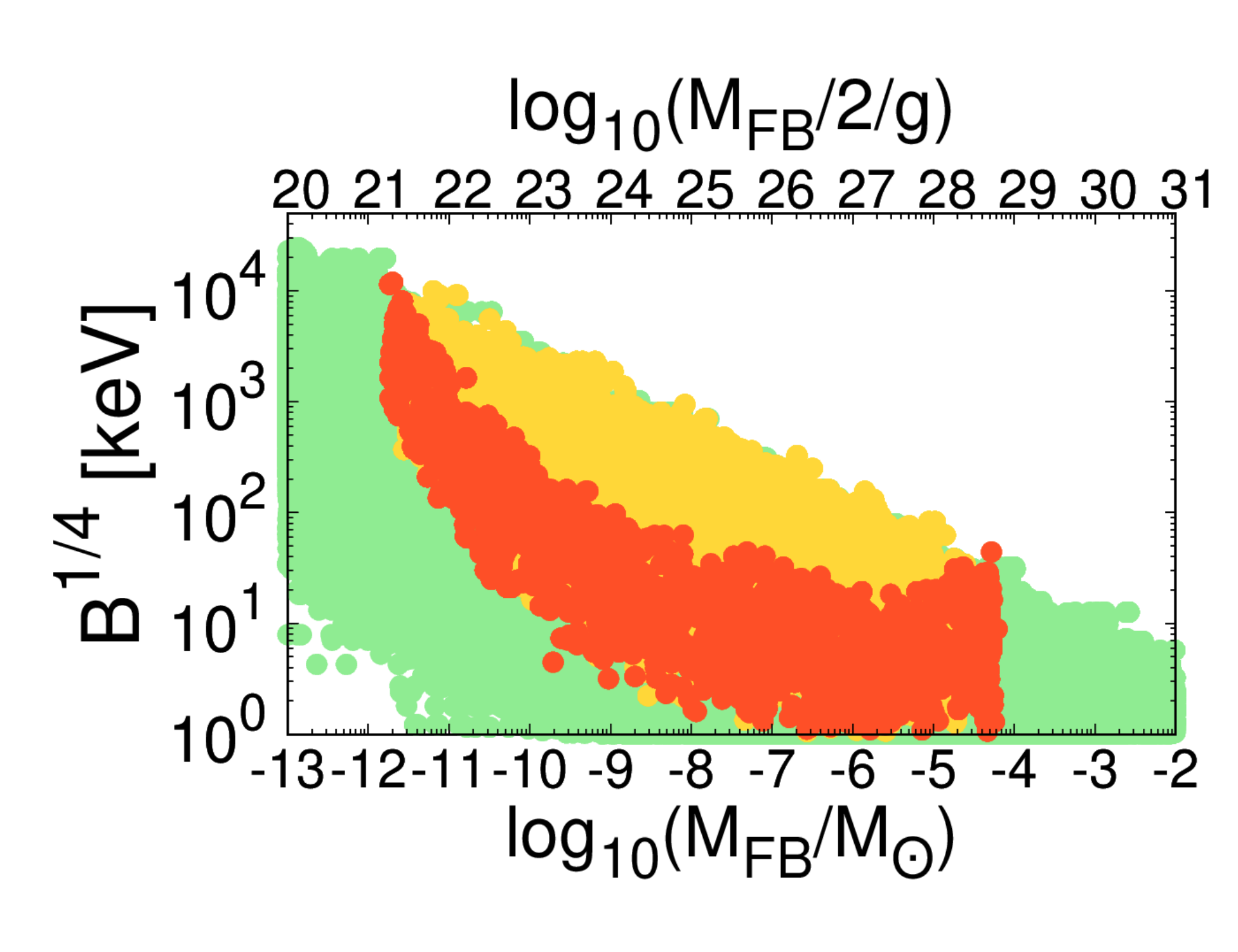}
      \includegraphics[height=1.1in,angle=0]{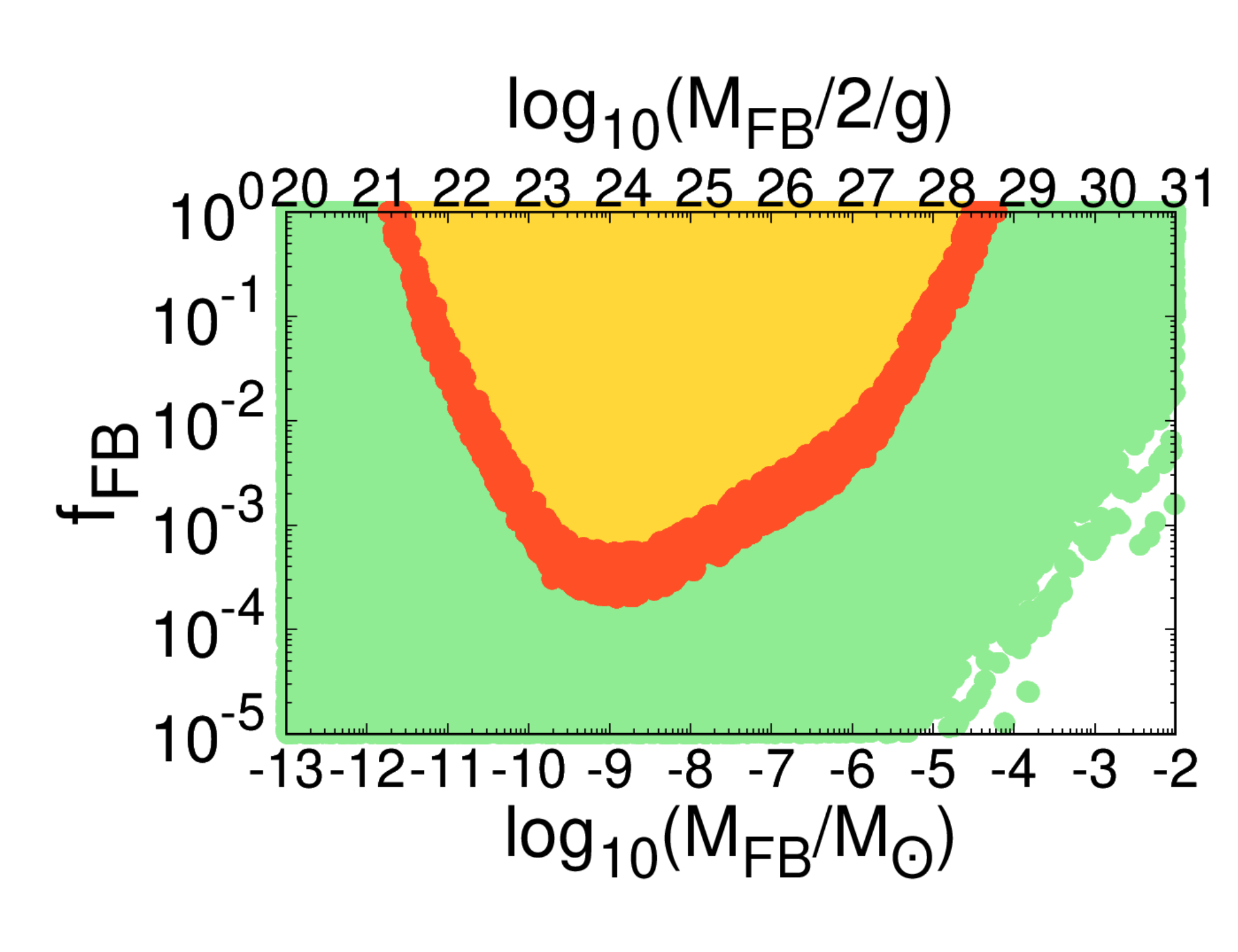}
      \caption{The parameter space excluded at 95\%~C.L. by Subaru-HSC's microlensing survey with 7 hours of observation is shown in yellow, and the expected 95\%~C.L. sensitivity with 70 hours of observation is shown in red.
      In the green regions, the FOPT generates GWs that are detectable by THEIA/$\mu$Ares. 
      The upper row is for extended mass distributions and the lower row is for monochromatic distributions. 
       $\langle M_{\rm FB} \rangle$ and $\langle R_{\rm FB} \rangle$ denote the average FB mass and radius, respectively. 
      }\label{fig:FB_microlensing}
\end{figure}

Figure~\ref{fig:FB_microlensing} shows the current 95\%~C.L. excluded region (yellow) and the projected sensitivity (red) in the FOPT parameter space based on $T_{\rm obs}=7$~hours and 70~hours, respectively. 
In the green regions, the FOPT generates GWs that are detectable by THEIA~\cite{Theia:2017xtk} or $\mu$Ares~\cite{Sesana:2019vho}.
The first (second) row shows results for extended (monochromatic) mass distributions.
Subaru-HSC is sensitive to FBs produced in FOPTs with 
$1~{\rm keV}\lesssim B^{1/4} \lesssim 10~{\rm MeV}$ and $10^{-10} \lesssim \eta_\chi \lesssim 10^{-4}$. 
The bottom-left edges in the regions arise because we have required $f_{\rm FB}\geq 10^{-5}$, and the bottom-right edge in the bottom-right panel corresponds to $B^{1/4}\geq 1~{\rm keV}$.
The rightmost panels show that the sensitivity to the fractional contribution of FBs to the dark matter density $f_{\rm FB}$ is maximum for FB masses $\sim 10^{-9}M_\odot$.
Outside this mass region the number of microlensing events is suppressed, either because the Einstein radius is too small (for lighter FBs) or the number density of FBs is too low (for heavier FBs). Note that Subaru-HSC has greater sensitivity to larger $\langle M_{\rm FB} \rangle$
for extended mass distributions than for the monochromatic case. This is understandable as the microlensing event rate receives large contributions from light FB masses in the extended mass distribution.
However, other two-dimensional projections do not differentiate between the extended and monochromatic mass distributions.
The inverse correlation between
the FB mass and the energy scale $B^{1/4}$ 
in the Subaru-HSC sensitivity for fixed values of $f_{\rm FB}$ is evident~\cite{Marfatia:2021twj} .

\subsection{Microlensing by primordial black holes}

PBHs produced by FB collapse inherit their mass distribution from FBs.
Since PBHs can be treated as point-like lenses, we repeat the microlensing analysis in the previous subsection 
in the point-like limit. Our results are obtained using geometric optics and neglect diffraction effects that become relevant if the Schwarzschild radius of the PBH is smaller than the wavelength of the microlensing survey. 
Keeping in mind that the Subaru-HSC survey is performed with optical wavelengths, PBHs lighter than about 
$10^{-10}M_\odot$ can not be studied~\cite{Sugiyama:2019dgt}. 

\begin{figure}[t]
      \centering
      \includegraphics[height=1.1in,angle=0]{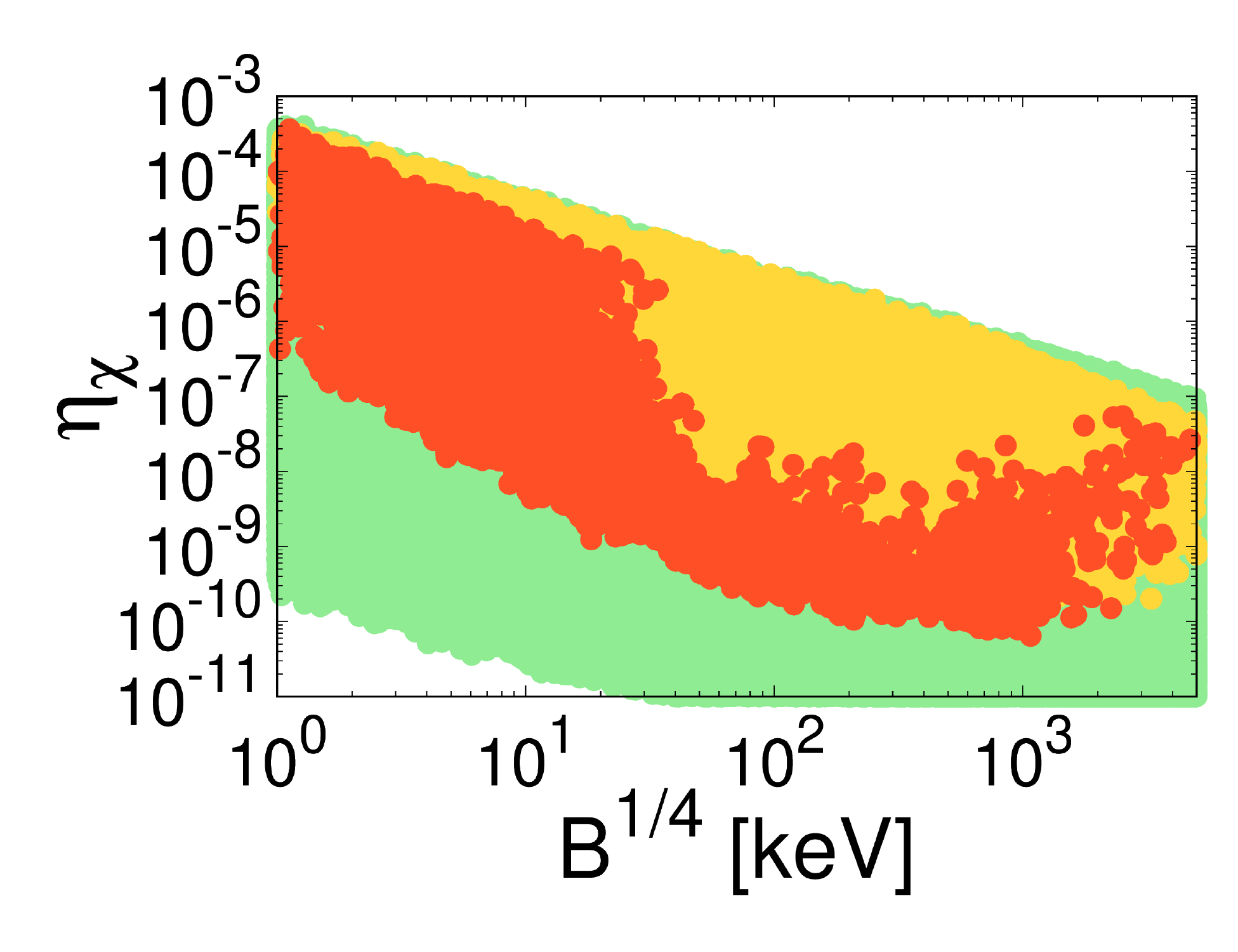}
      \includegraphics[height=1.07in,angle=0]{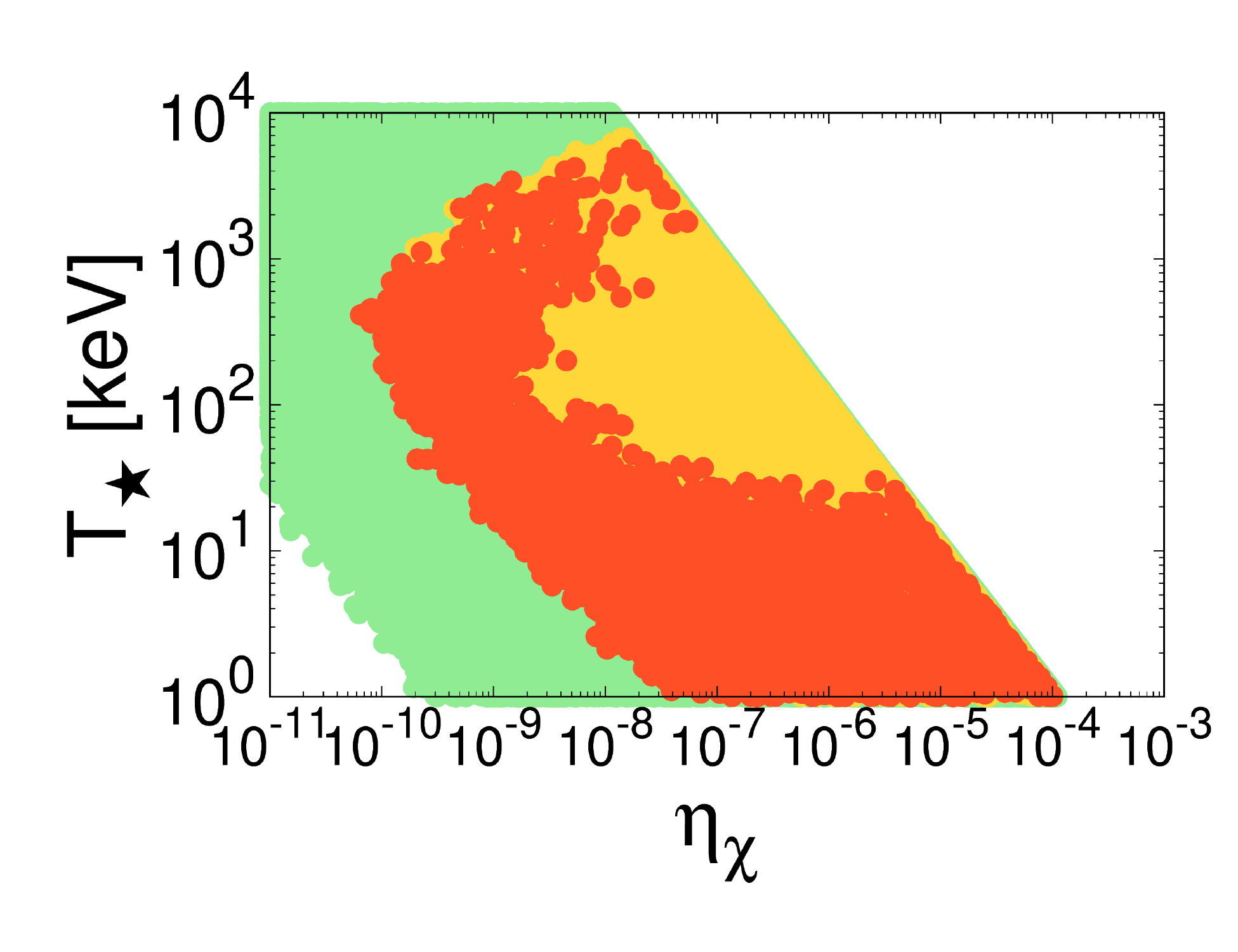}
      \includegraphics[height=1.1in,angle=0]{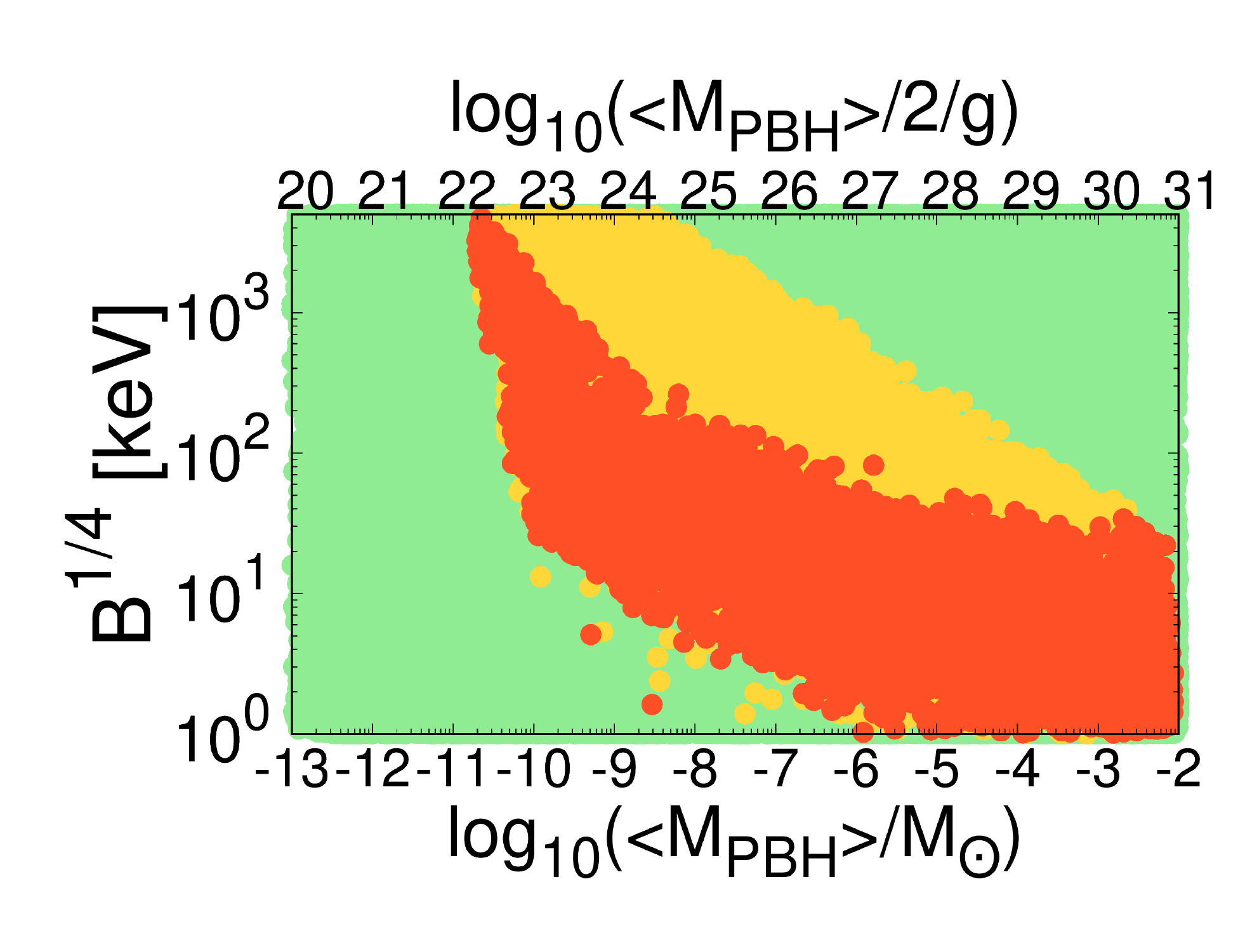}
      \includegraphics[height=1.1in,angle=0]{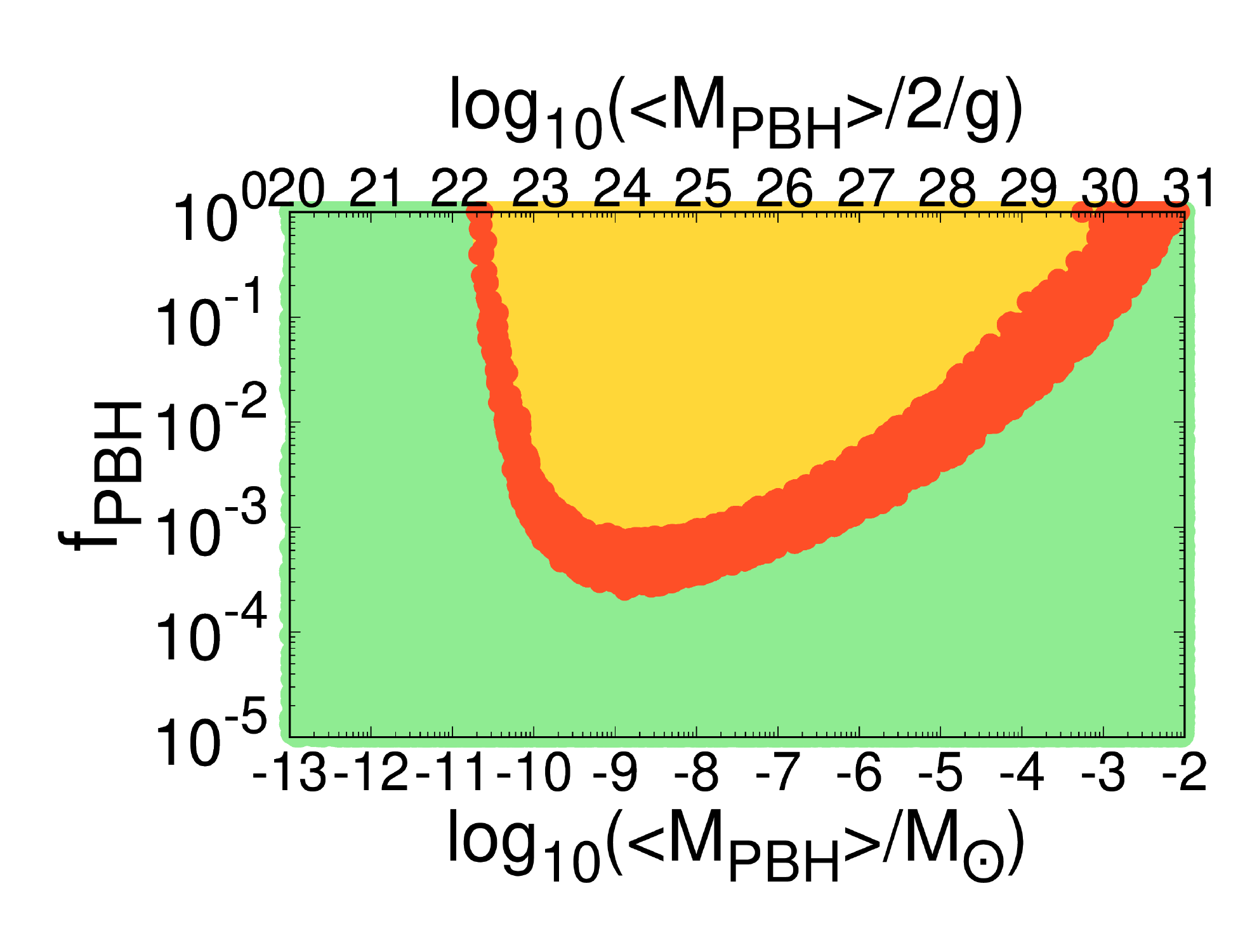}
      \includegraphics[height=1.1in,angle=0]{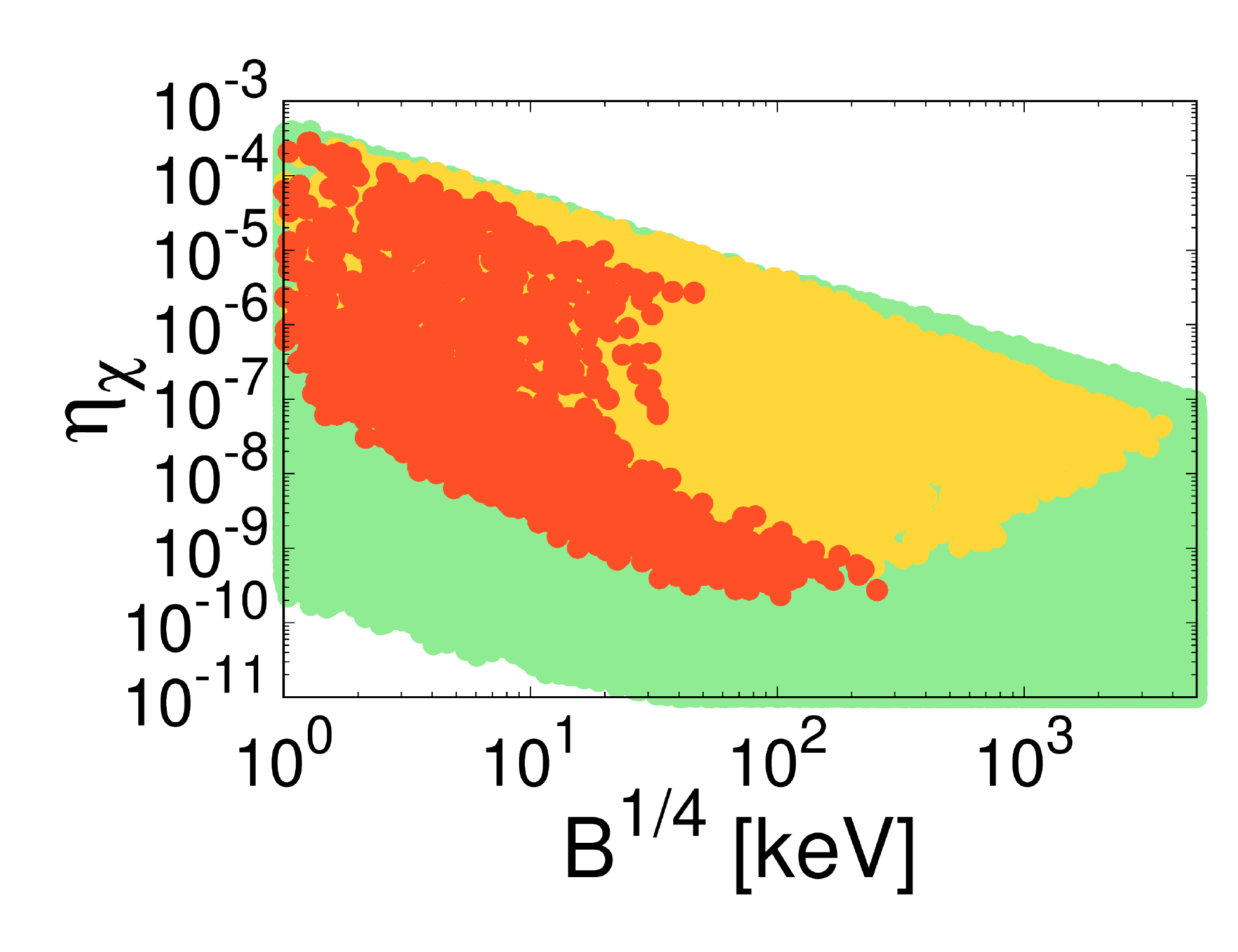}
      \includegraphics[height=1.07in,angle=0]{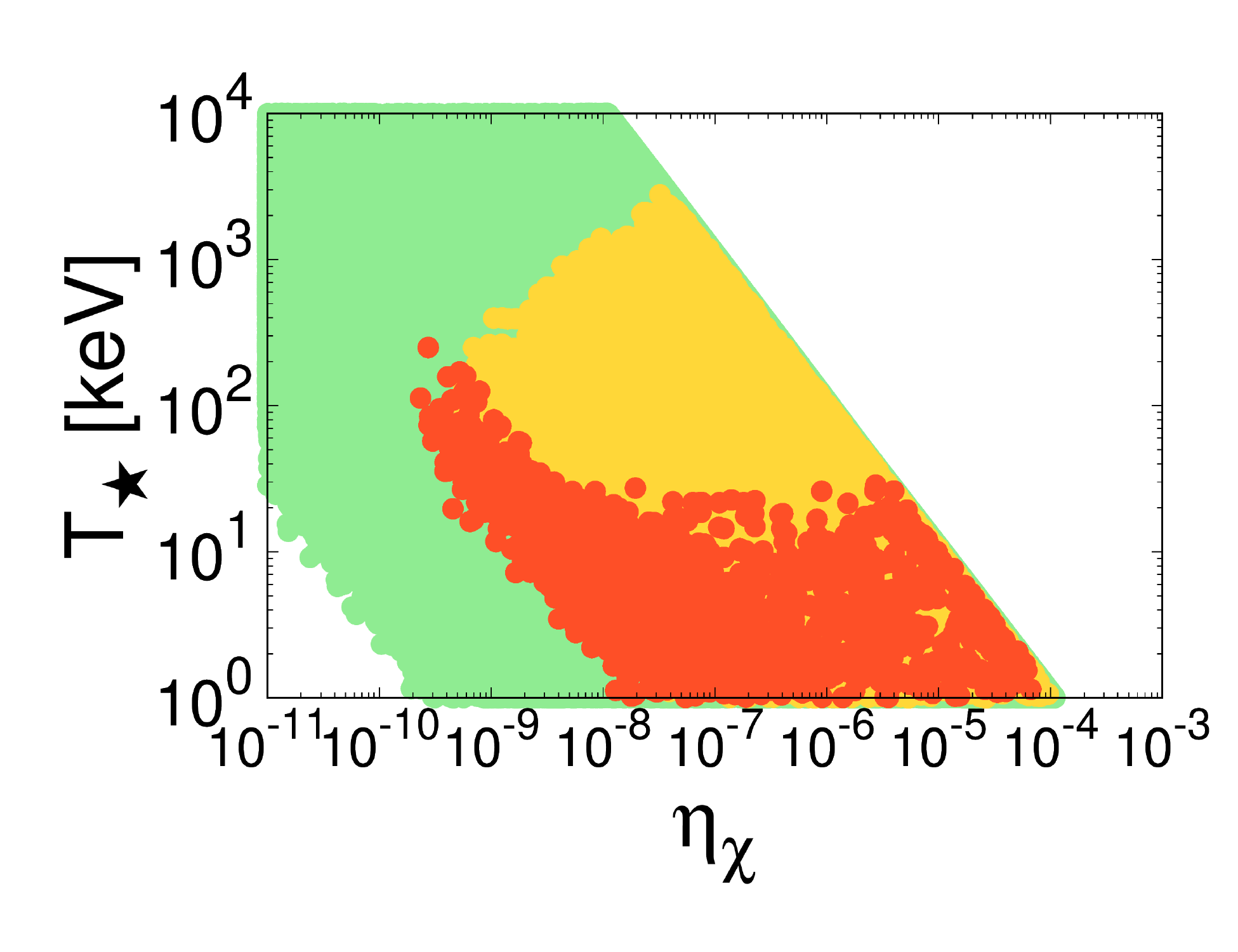}
      \includegraphics[height=1.1in,angle=0]{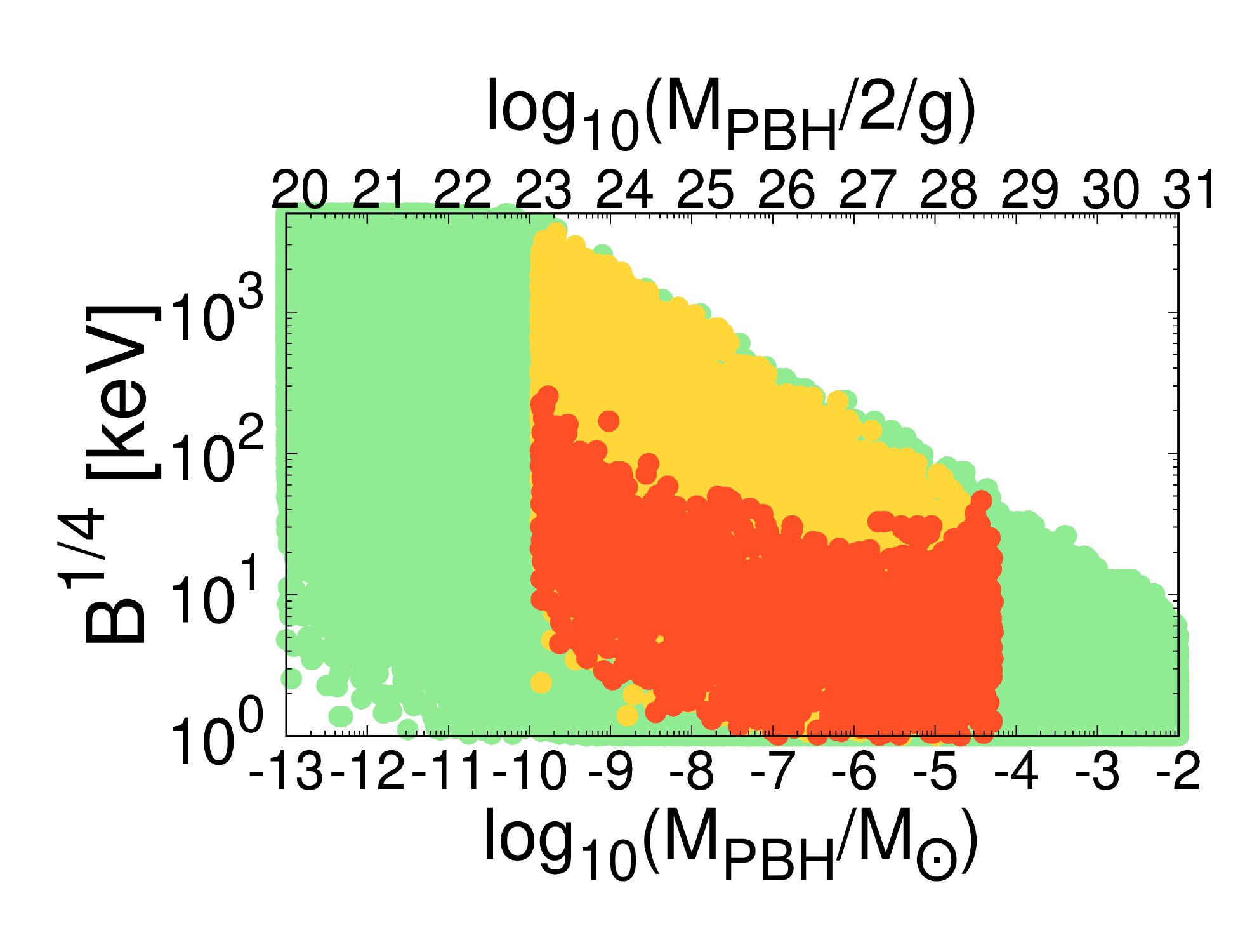}
      \includegraphics[height=1.1in,angle=0]{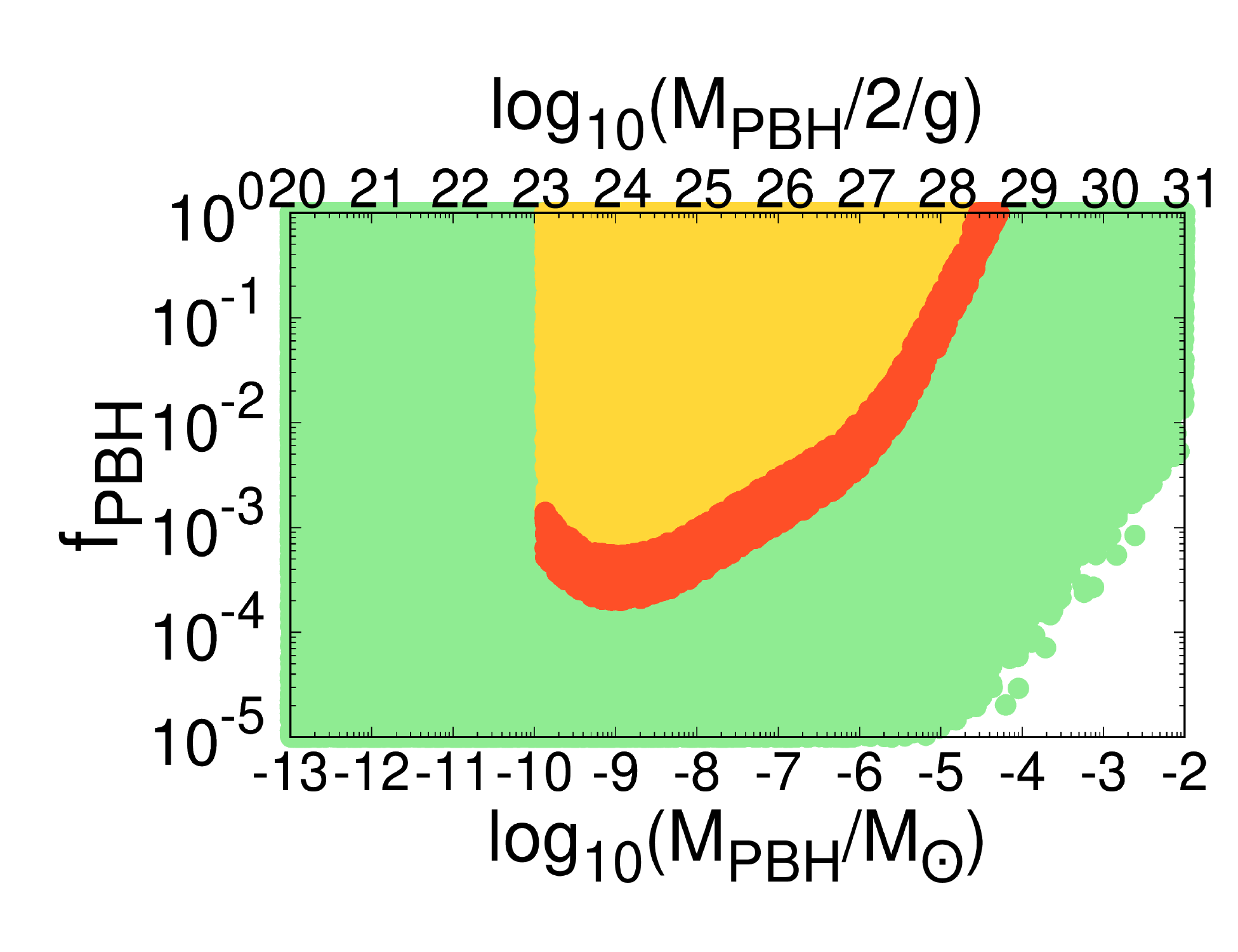}
      \caption{
      The same as Fig.~\ref{fig:FB_microlensing}, but for microlensing by PBHs.
      }\label{fig:PBH_microlensing}
\end{figure}

Figure~\ref{fig:PBH_microlensing} is similar to Fig.~\ref{fig:FB_microlensing} for PBHs with an extended (first row) and monochromatic (second row) mass distributions. 
For the latter case, the PBH mass is restricted to $10^{-10}\lesssim M_{\rm PBH}/M_\odot \lesssim 10^{-4}$. On the other hand, the sensitivity region is widened for extended mass distributions. This slightly modifies correlations between the FOPT parameters. For example, from the $(B^{1/4},\eta_\chi)$ panels, it is evident that $B^{1/4}\gtrsim 100~{\rm keV}$ becomes relevant to Subaru-HSC for $T_{\rm obs}=70$~hours (red regions) if PBHs have extended distributions.
This corresponds to greater sensitivity to larger $\langle M_{\rm PBH} \rangle$ (in the top-right panel) from lighter PBHs in the mass distribution.

\bigskip

\section{Extragalactic gamma-ray spectrum}
\label{sec:gamma_ray}

The PBH mass distribution may be imprinted in the extragalactic gamma-ray background (EGB) through Hawking radiation.
The full-sky extragalactic photon flux for an extended PBH mass distribution is~\cite{Tseng:2022jta}\footnote{The expression in Ref.~\cite{Tseng:2022jta} has an extra factor of $1/4\pi$ in the normalization.} 
\begin{eqnarray}
\label{eq:extra_photon}
 \frac{d\Phi}{dE}&=& \int^{\mathrm{min}(t_{\mathrm{eva}},t_0)}_{t_\mathrm{CMB}}dt~
    {c[1+z(t)]\,n_{\rm PBH}|_0
    \int dM_{\rm PBH} \frac{1}{{n_{\rm PBH}}\vert_\star} \frac{d{n_{\rm PBH}}\vert_\star}{dM_{\rm PBH}} \frac{d^2N_\gamma(M_{\rm PBH})}{d\tilde{E}dt}\bigg|_{\tilde{E}=[1+z(t)]E}}\,, \nonumber \\
    \end{eqnarray}
where $n_{\rm PBH}|_\star$ is the PBH number density at percolation, and $n_{\rm PBH}|_0$ is the current number density, or for short-lived PBHs the number density they would have today had they not evaporated. Similarly, the  fractional contribution
of PBHs to the dark matter density $f_{\rm PBH}$ is interpreted as the value today had they not evaporated away. Photons emitted after the formation of the cosmic microwave background (CMB) until the PBH completely evaporates ($t_{\rm eva}$) or that are being emitted today ($t_0$) contribute to the flux.
We calculate the flux from Hawking evaporation using BlackHawk v2.1, which 
accounts for mass evolution of an extended PBH mass distribution~\cite{Arbey:2019mbc,Arbey:2021mbl}.
Note that the photon energy observed today $E$ is 
redshifted from the energy at emission from the PBH $\tilde{E}$. 

\begin{figure}[t]
        \centering
        {\includegraphics[width=10cm]{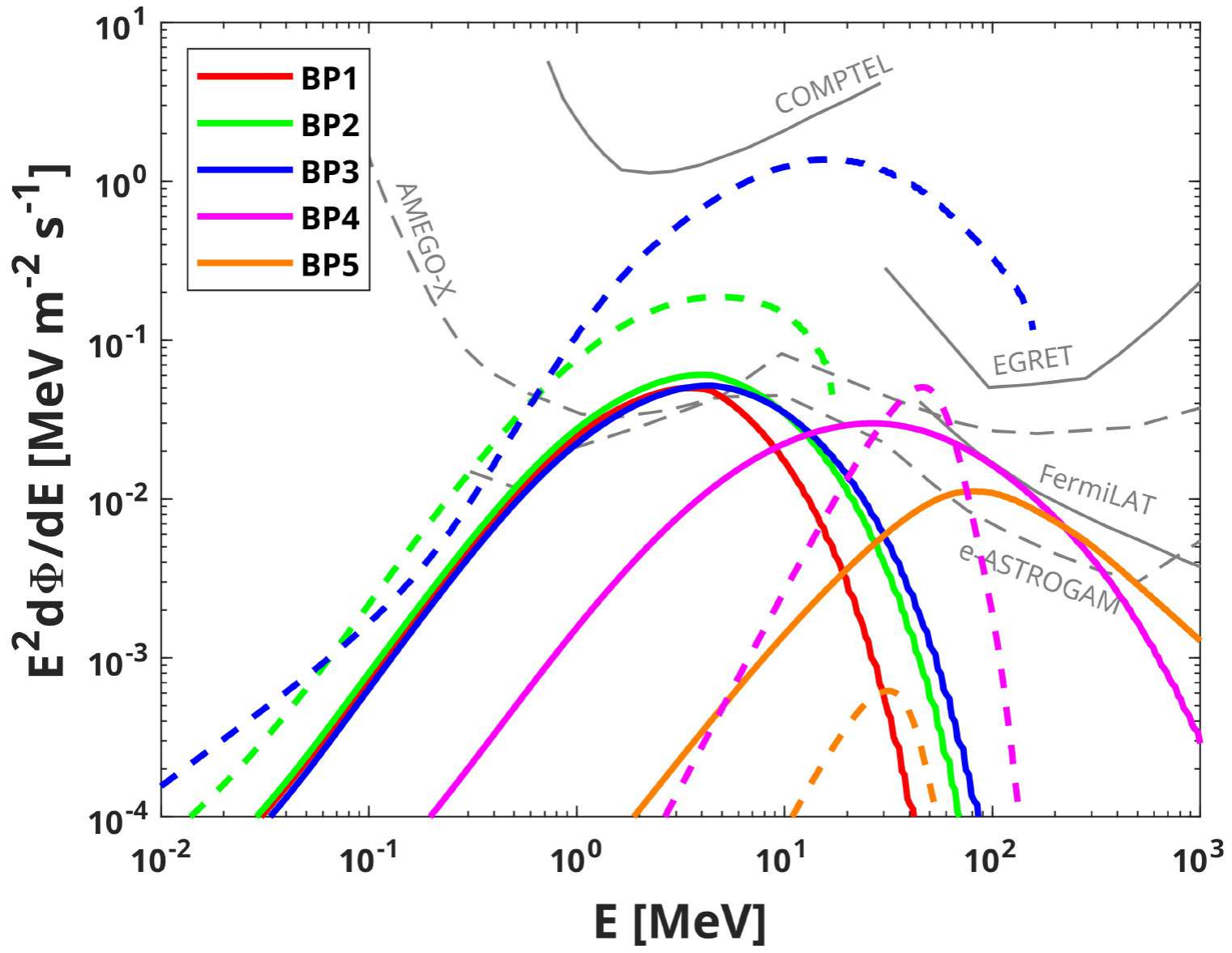}}
        \caption{Extragalactic photon spectra from PBH evaporation for the benchmark points in Table~\ref{table1}. The solid (dashed) curves correspond to extended (monochromatic) mass distributions. 
         The solid curves for {\bf BP-1} and {\bf BP-2} exhibit a mild spectral break at about 5~MeV. While all the BPs give $>3\sigma$ signals at AMEGO-X/e-ASTROGAM for extended mass distributions, the monochromatic distribution for {\bf BP-3} is excluded by current data. There is no signal for the monochromatic distribution for {\bf BP-1} because PBHs lighter than $1.38\times 10^{-20}M_\odot$ evaporate before the epoch of last scattering.      }\label{fig:g}
\end{figure}

The EGB spectra for {\bf BP-1} to {\bf BP-5}, which produce PBHs, are shown in Fig.~\ref{fig:g}. Some of the spectra for monochromatic distributions (dashed curves) end abruptly because BlackHawk restricts the maximum  energy of Hawking radiation to be $\tilde{E}=5$~GeV, and as the PBH evaporates, its temperature exceeds this limit. The cutoffs are at different energies because the redshifting down to $E$ depends on the evolution of the PBH. 

Figure~\ref{fig:g} shows a mild break in the spectra for {\bf BP-1} and {\bf BP-2} which have extended mass distributions (solid curves) dominated by PBHs lighter than $7.28 \times 10^{14}$~g ($\sim 4\times 10^{-19} M_\odot$) that have evaporated before today. This is evident from the rapidly falling mass distributions for those BPs in Fig.~\ref{figmd}. 
The spectral break becomes smoother from {\bf BP-1} to {\bf BP-3} because the cliffs in the mass distributions shift to the right. Also note that the amplitude of the EGB spectrum for the monochromatic distribution for {\bf BP-3} is much larger than that for an extended distribution.
There is no signal for the monochromatic distribution for {\bf BP-1} because PBHs evaporated before the epoch of last scattering. 
The mild spectral break between $5 - 10$~MeV for light PBHs facilitates differentiation between an extended mass distribution and a monochromatic distribution for the same set of FOPT input parameters. Note that points that show a spectral break in the EGB, do not produce a detectable GW signal.

{\bf BP-5} with a broad PBH mass distribution has a spectral peak at $\sim 100$~MeV due to contributions from PBHs with lifetimes longer than the age of the Universe. The corresponding spectrum for a monochromatic distribution is softer and has a smaller amplitude because the monochromatic PBH masses are too large to produce significant Hawking emission.

{\bf BP-4} illustrates a scenario intermediate to the preceding two in that the fall-off in the PBH mass distribution occurs closer to $M_{\rm PBH}\simeq 7\times 10^{14}$~g. Consequently, the amplitudes of the extragalactic gamma-ray spectra  for the extended and monochromatic mass distributions are comparable and the spectrum for the extended distribution does not exhibit a spectral break.

\begin{figure}[t]
      \centering
      \includegraphics[height=1.05in,angle=0]{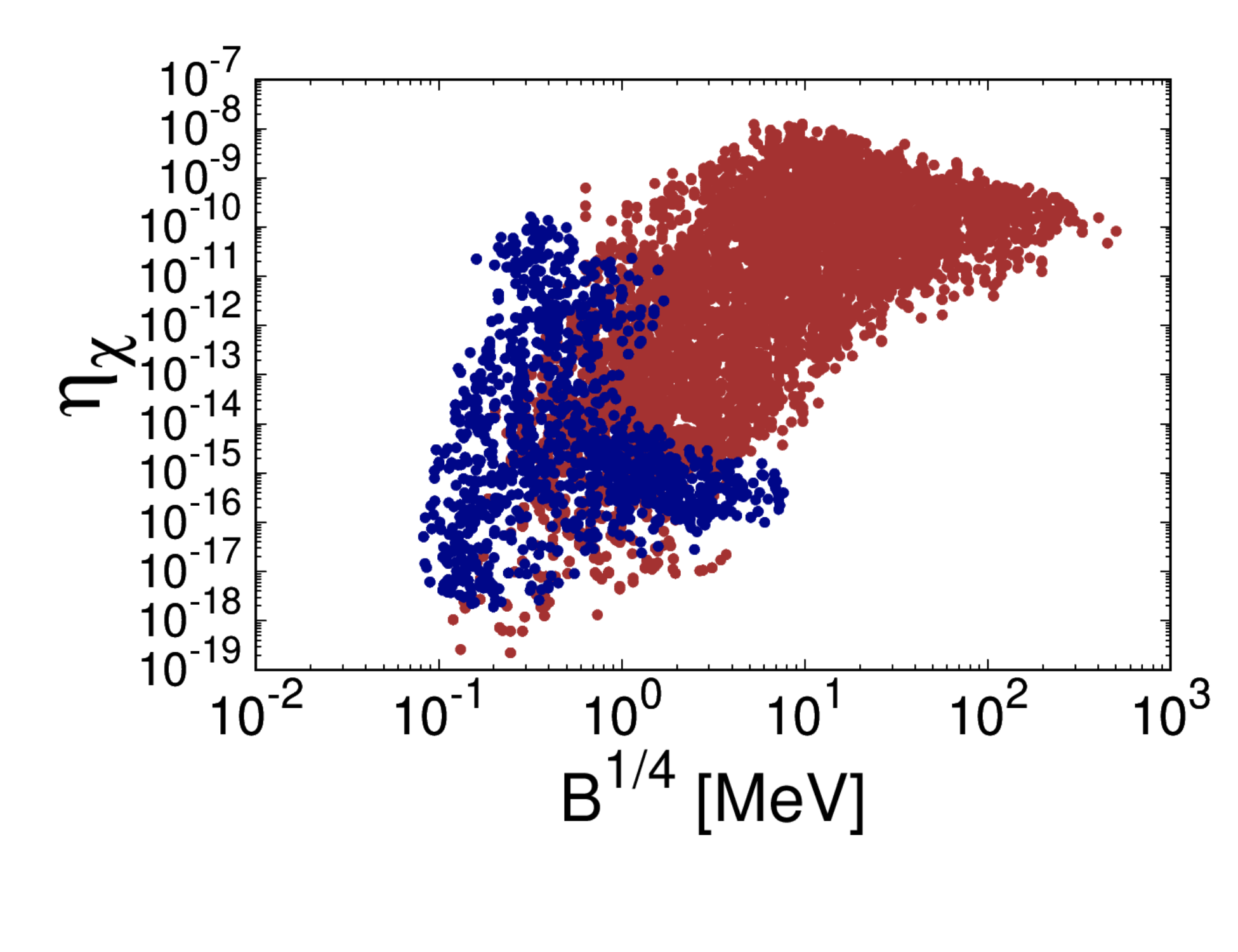}
      \includegraphics[height=1.07in,angle=0]{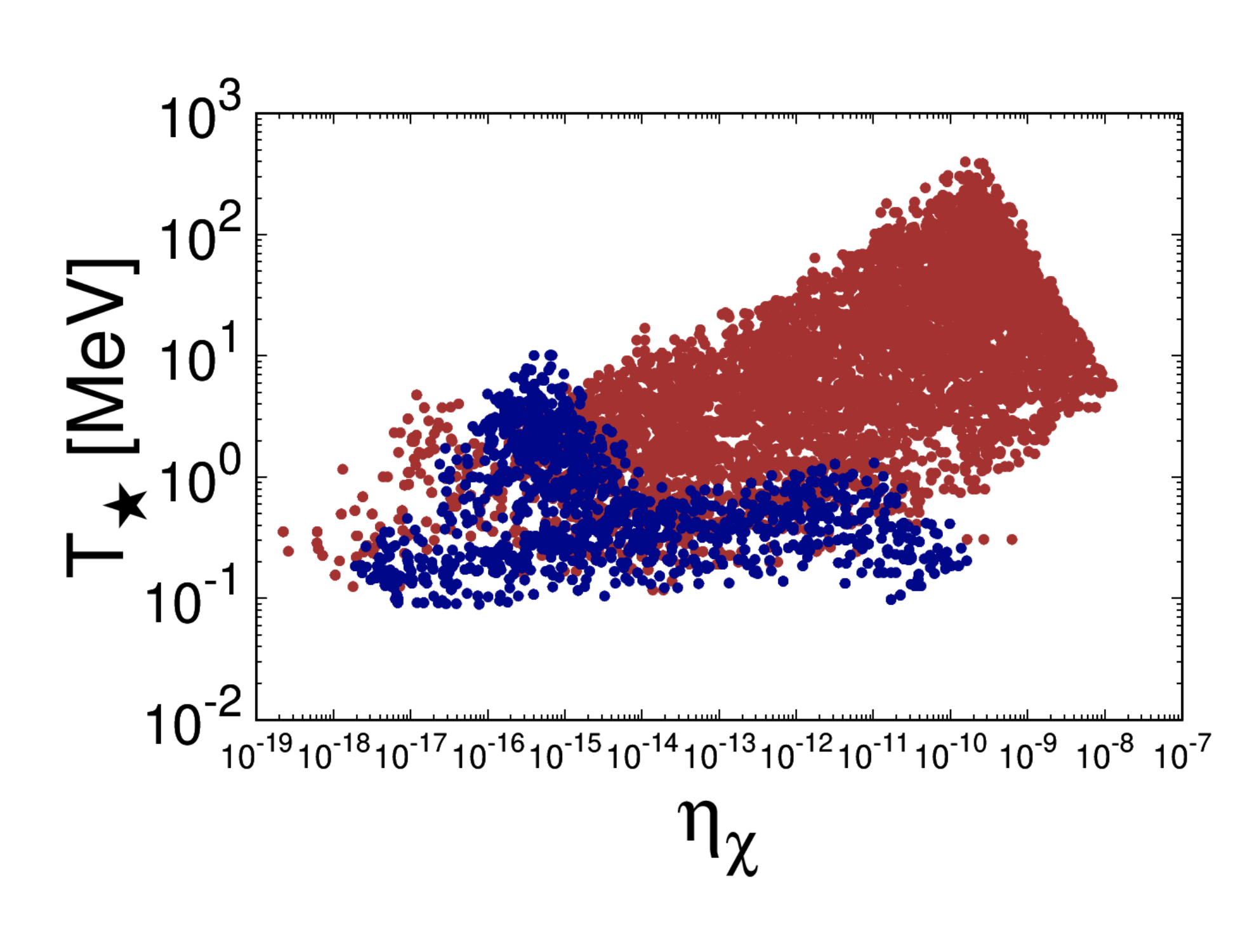}
      \includegraphics[height=1.1
      in,angle=0]{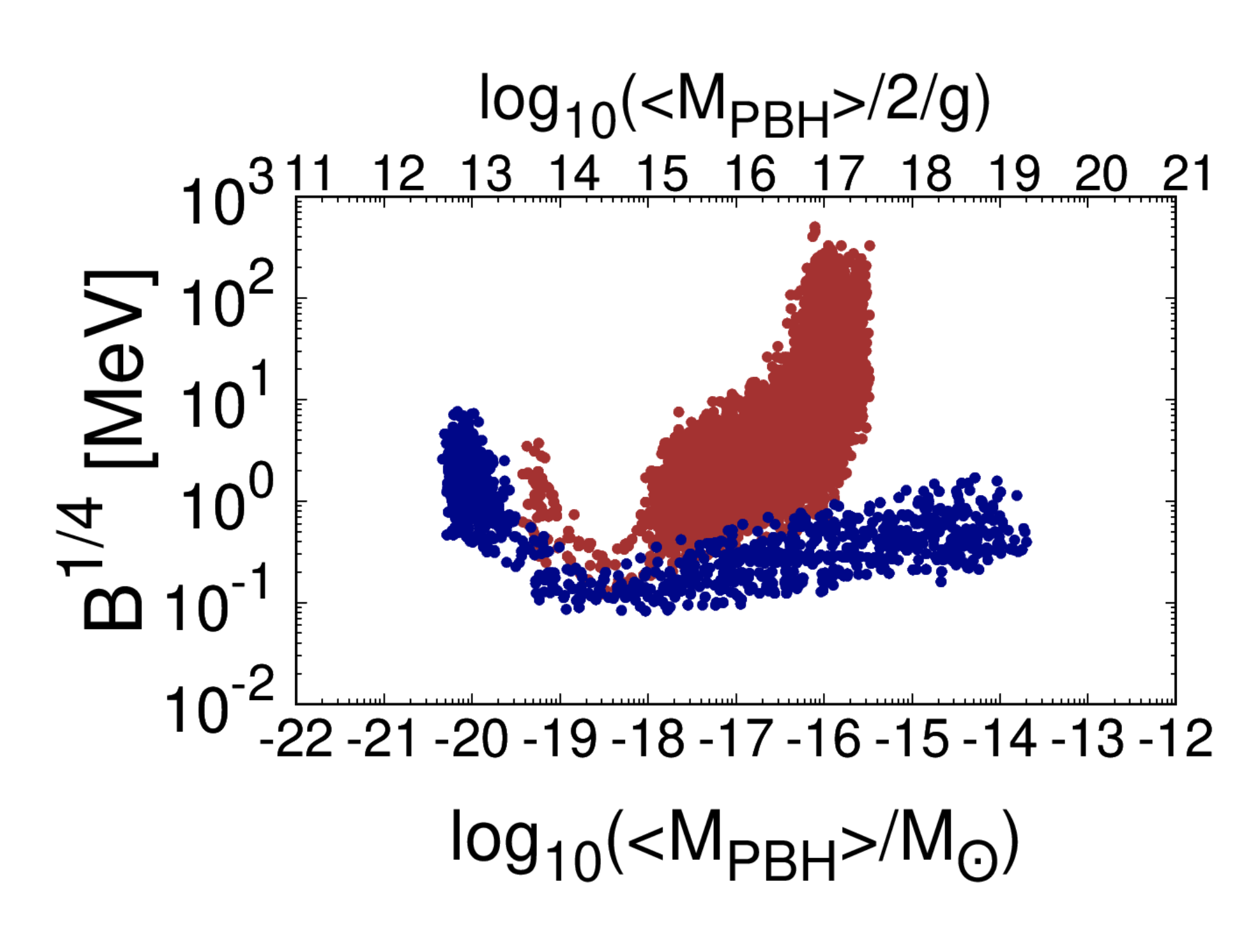}
      \includegraphics[height=1.15in,angle=0]{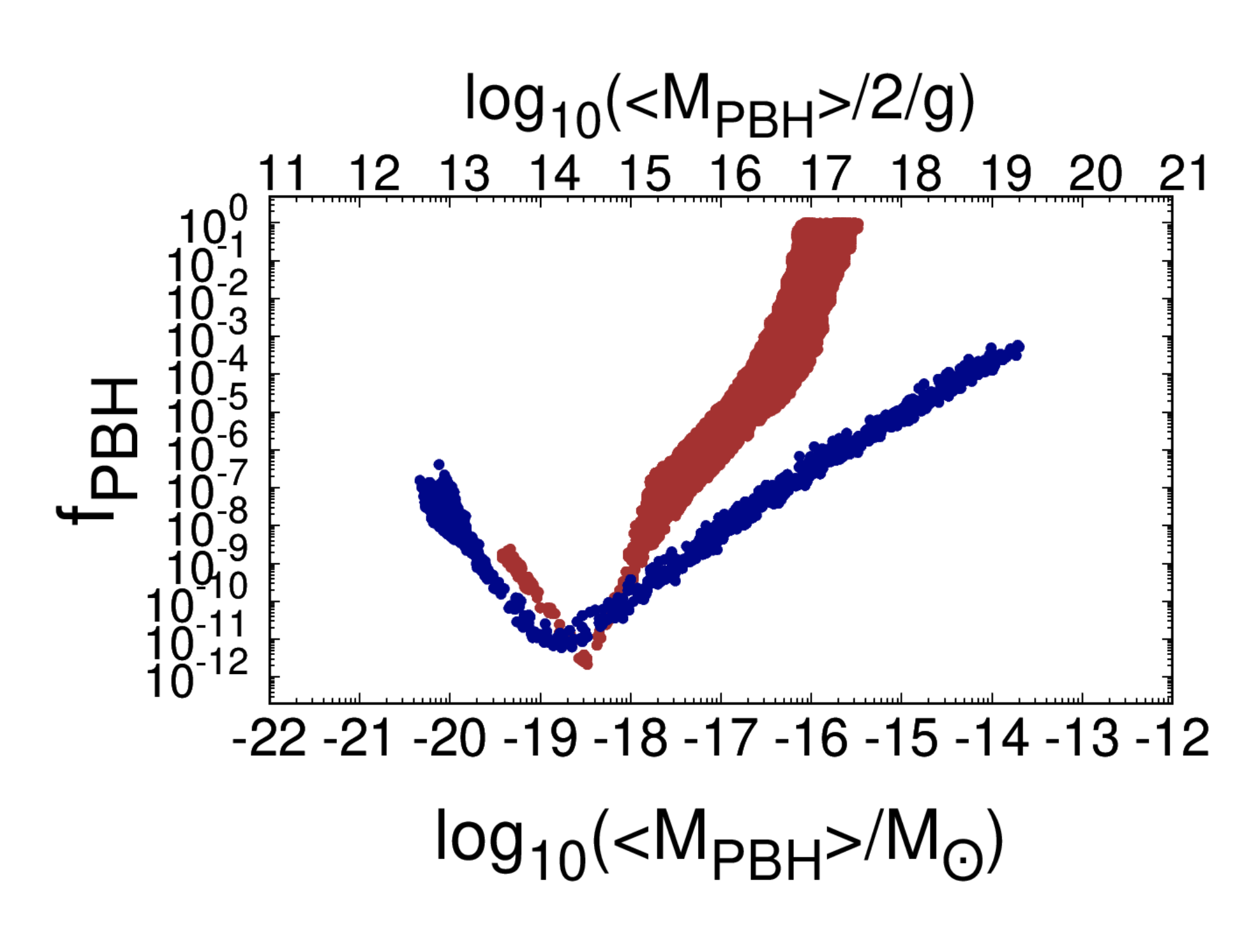}     
      \caption{The regions of parameter space that produce a $3\sigma$ signal of a diffuse extragalactic X-ray/gamma-ray background at AMEGO-X and e-ASTROGAM from PBH evaporation. 
      The blue (red) points correspond to extended (monochromatic) mass distributions.     
      }\label{fig1}
\end{figure}

All points in Fig.~\ref{fig1} are compatible with the $3\sigma$ upper limits from COMPTEL~\cite{COMPTEL_v2} and Fermi-LAT~\cite{Fermi-LAT:2018pfs} and produce a $3\sigma$ signal at AMEGO-X~\cite{Fleischhack:2021mhc} and e-ASTROGAM~\cite{e-ASTROGAM:2017pxr}.
They are also compatible with big bang nucleosynthesis, CMB, and EGB constraints on the fraction of the energy density of the Universe in PBHs at formation, $f_\phi\equiv\rho_{\rm PBH}(T_\phi)/\rho(T_\phi)$. We apply the bounds in Fig.~4 of Ref.~\cite{Carr:2020gox}, which although derived for monochromatic mass distributions, can be approximately applied to extended distributions because they increase monotonically in wide ranges of PBH mass~\cite{Dent:2025lwe}; we do not consider these bounds in the ranges in which they are not monotonically increasing.
For an extended distribution, we integrate the PBH number density in 
Eq.~(\ref{dist}) up to $M_{\rm cut}$, then assuming that all the PBHs have mass $M_{\rm cut}$,  we calculate $f_\phi|_{M_{\rm cut}}$ following Ref.~\cite{Marfatia:2021hcp}, which is an overestimate of $f_\phi$. Then we vary $M_{\rm cut}$ and require that every value of $f_\phi|_{M_{\rm cut}}$ satisfies each of the big bang nucleosynthesis, CMB, and EGB upper limits in Ref.~\cite{Carr:2020gox}. 

From Fig.~\ref{fig1}, the energy scale of the phase transition, 
$0.1~{\rm MeV}\, \lsim \, B^{1/4} \, \lsim \, 10~{\rm MeV}$, is correlated with the average PBH mass $10^{13}~{\rm g}\, \lsim \langle M_{\rm PBH} \rangle \lsim \, 10^{20}~{\rm g}$. 
The lowest fractional abundance $f_{\rm PBH} \simeq 10^{-12}$ occurs for $\langle M_{\rm PBH} \rangle \simeq 7\times 10^{14}~{\rm g}$ for which the PBH lifetime coincides with the age of the Universe. For $\langle M_{\rm PBH} \rangle \lesssim 7\times 10^{14}~{\rm g}$, lighter PBHs evaporated before today so that larger values of $f_{\rm PBH}$ are required to generate a substantial gamma-ray flux.
Conversely, $f_{\rm PBH}$ increases for points with $\langle M_{\rm PBH} \rangle \gsim 7\times 10^{14}~{\rm g}$  to compensate for the suppressed Hawking emission. 
The blue points (corresponding to extended mass distributions), with large average PBH masses $\langle M_{\rm PBH} \rangle \gsim 10^{14}$~g, have $B^{1/4}, T_\star \lsim 1$~MeV.

\bigskip

\section{Summary}
\label{sec:summary}

Extended vacuum bubble radius distributions lead to extended mass distributions for FBs and PBHs produced in the dark FOPT. 
We studied phenomenological implications of the extended distributions for GW signals, microlensing signals of FBs and PBHs, and Hawking evaporation signals of PBHs. 

We applied the double broken power law GW spectrum, which uses an average TV bubble radius, to model the GW spectrum from extended TV bubble radius distributions. We find the spectral peaks shift to frequencies lower than for the average bubble radius, and spectral broadening is significant below the peak frequency.
The TV bubble radius distribution qualitatively impacts the GW spectrum and presents another important uncertainty in its calculation.

Optical microlensing surveys like Subaru-HSC are not sensitive to PBHs lighter than about $10^{-10} M_\odot$ (associated with FOPT energy scales $B^{1/4} \gsim 10$~keV) because their Schwarzschild radius is smaller than the optical wavelength, which renders the use of geometric optics invalid. However, such surveys can probe extended PBH mass distributions with {\it average} PBH masses below $10^{-10} M_\odot$ and correspondingly FOPTs with $B^{1/4} \gsim 10$~keV.
Needless to say, individual microlensing events can not differentiate between an extended mass distribution and a monochromatic one.

We also showed that the PBH mass distribution can imprint itself in the extragalactic gamma-ray spectrum from Hawking evaporation.
Specifically, a mild spectral break is a feature that differentiates an extended mass distribution from a monochromatic distribution. This signal is detectable
at AMEGO-X or e-ASTROGAM.

\bigskip

\section*{Acknowledgments}
We thank J.~Kumar for discussions. D.M. is supported in part by the U.S. Department of Energy under Grant No.~de-sc0010504. 
P.Y.T. is supported in part by the National Science and Technology Council under
Grant No. NSTC-111-2112-M-007-012-MY3, and Physics Division of the National Center for Theoretical Sciences of Taiwan with Grant NSTC 114-2124-M-002-003.
Y.M.Y. is supported in part by the Ministry of Education under Grant No. 111J0382I4.

\medskip

%\newpage
%%%%%%%%%%%%%%%%%%%%%%%%%%%%%%%
%%%%%%%%%%% Appendix %%%%%%%%%%%
%%%%%%%%%%%%%%%%%%%%%%%%%%%%%%%

%\newpage

\end{document}